\documentclass[camera]{jpaper}
\usepackage[nocompress]{cite}
\usepackage[redeflists]{IEEEtrantools}

\usepackage{soul}

\usepackage{amsmath}
\usepackage{graphicx}
\usepackage[linesnumbered,ruled]{algorithm2e}

\usepackage{algpseudocode}
\usepackage{titlesec}
\algrenewcommand\algorithmiccomment[2][\normalsize]{{#1\hfill\(\triangleright\) #2}}
\usepackage{geometry} \titlespacing*{\section}{0pt}{3pt}{-1pt}
\titlespacing*{\subsection}{0pt}{3pt}{1pt}
\titlespacing*{\subsubsection}{0pt}{1pt}{0pt}

\usepackage{array}
\makeatletter
\let\MYcaption\@makecaption
\makeatother

\usepackage[font=footnotesize]{subcaption}

\makeatletter
\let\@makecaption\MYcaption
\makeatother

\usepackage{fixltx2e}

\usepackage{dblfloatfix}
\usepackage[nolessnomore, italic]{mathastext}
\usepackage[T1]{fontenc}
\usepackage[usenames,dvipsnames,svgnames,table]{xcolor}
\usepackage{multirow}
\usepackage{hhline}
\usepackage[normalem]{ulem}
\usepackage{setspace}
\usepackage{indentfirst}
\usepackage{footmisc}
\usepackage{datetime}

\usepackage{pifont}

\newif\ifcameraready
\camerareadytrue

\ifcameraready
\newcommand{\affilIntel}[0]{\textsuperscript{$\star$}}

\newcommand{\affilETH}[0]{\textsuperscript{\S}}
\newcommand{\affilUOT}[0]{\textsuperscript{$\dagger$}}

\else
\newcommand{\affilIntel}[0]{\textsuperscript{$\ddagger$}}

\newcommand{\affilETH}[0]{\textsuperscript{\S}}
\fi

\newcommand{\ch}[1]{{\color{black}#1}}
\definecolor{amber}{rgb}{1.0, 0.49, 0.0}
\definecolor{darkgreen}{rgb}{0.0, 0.2, 0.13}
\definecolor{darkbyzantium}{rgb}{0.36, 0.22, 0.33}
\definecolor{darkseagreen}{rgb}{0.56, 0.74, 0.56}
\definecolor{darkspringgreen}{rgb}{0.09, 0.45, 0.27}
\definecolor{dollarbill}{rgb}{0.52, 0.73, 0.4}

\newcommand{\agy}[1]{{\color{black}#1}}

\newcommand{\hj}[1]{{\color{black} #1}}
\newcommand{\hjg}[1]{{\color{black}#1}}
\newcommand{\hjk}[1]{{\color{black}#1}}
\newcommand{\hjl}[1]{{\color{black}#1}}
\newcommand{\hjm}[1]{{\color{black}#1}}
\newcommand{\hjn}[1]{{\color{black}#1}}
\newcommand{\nv}[1]{{\color{black}#1}}

\fancypagestyle{firstpage}{
 
 \fancyhead[C]{\normalsize{HPCA 2019 Submission
 \textbf{\#\microsubmissionnumber} -- Confidential Draft -- Do NOT Distribute!! \\
 }}
 \pagenumbering{arabic}
}

\begin{document}
\bstctlcite{IEEEexample:BSTcontrol} 


\title{\vspace{-9pt}SysScale: Exploiting Multi-domain Dynamic Voltage and Frequency Scaling for   Energy Efficient  Mobile Processors\vspace{3pt}}


%


\author{
{Jawad Haj-Yahya\affilETH}\qquad%
{Mohammed Alser\affilETH}\qquad%
{Jeremie Kim\affilETH}\qquad 
{A. Giray Ya\u{g}l{\i}k\c{c}{\i}\affilETH}\qquad \\
{Nandita Vijaykumar \affilETH\affilIntel\affilUOT}\qquad%
{Efraim Rotem\affilIntel}\qquad%
\vspace{6pt}
{Onur Mutlu\affilETH}\qquad\\%
\emph{{\affilETH ETH Z{\"u}rich \qquad   \affilIntel Intel  \qquad \affilUOT University of Toronto}}%
\vspace{-5pt}%
}

\newcommand{\ballnumber}[1]{\tikz[baseline=(myanchor.base)] \node[circle,fill=black,inner sep=1pt] (myanchor) {\color{-.}\bfseries\footnotesize #1};}

\newcommand*\circled[1]{\tikz[baseline=(char.base)]{
            \node[shape=circle,draw,inner sep=1pt] (char) {#1};}}
\newcommand*\circledb[1]{\tikz[baseline=(char.base)]{
            \node[shape=circle,fill,inner sep=0.5pt] (char) {\textcolor{white}{#1}};}}            
            
\def\BibTeX{{\rm B\kern-.05em{\sc i\kern-.025em b}\kern-.08em
    T\kern-.1667em\lower.7ex\hbox{E}\kern-.125emX}}

\newcommand{\tech}{SysScale\xspace}


\maketitle
\thispagestyle{plain} 
\pagestyle{plain}

\setstretch{0.978}
\renewcommand{\footnotelayout}{\setstretch{0.9}}



\fancyhead{}
\thispagestyle{plain}
\pagestyle{plain}
\fancypagestyle{firststyle}
{
  \fancyfoot[C]{\thepage}
}
\thispagestyle{firststyle}
\pagestyle{firststyle}




%


\begin{abstract}



There are three domains in \agy{a} modern thermally-constrained mobile system-on-chip (SoC): compute, IO, and memory.
We observe that a modern SoC typically allocates  a fixed power budget, corresponding to worst-case performance demands, to the IO and memory domains even if they are underutilized.
The resulting unfair allocation of the power budget across domains can cause two major issues: 
1) the IO and memory domains can operate at a higher frequency and voltage than necessary, \agy{increasing} power consumption and 2) the unused power budget of the IO and memory domains cannot be used to increase the throughput of the compute domain\agy{, hampering performance.}
To avoid these issues, it is crucial to dynamically orchestrate the distribution of the SoC power budget across the three domains based on their actual performance demands.
We propose \emph{\tech}, a new multi-domain power management \agy{technique} to improve the energy efficiency of mobile SoCs.
{\tech} is based on three key ideas. 
First, {\tech} introduces an \agy{accurate algorithm} to predict the performance (e.g., bandwidth and latency) demands of the three SoC domains.
Second, {\tech} uses a new DVFS (dynamic voltage and frequency scaling) mechanism to distribute the SoC power to each domain according to the predicted performance demands.  This mechanism is designed to minimize the significant latency overheads associated with applying DVFS across multiple domains. 
Third, in addition to using a global DVFS mechanism, {\tech} uses domain-specialized techniques to optimize the energy efficiency of \emph{each} domain at different operating points.
We implement {\tech} on an Intel Skylake microprocessor for mobile devices and evaluate it using a wide variety of SPEC CPU2006, graphics (3DMark), and battery life workloads (e.g., video playback). 
On a 2-core Skylake,
{\tech} improves the performance of SPEC CPU2006 and 3DMark workloads by up to $16\%$ and $8.9\%$ ($9.2\%$ and $7.9\%$ on average), respectively.
For battery life workloads, which typically have fixed performance demands, {\tech} reduces the average power consumption by up to $10.7\%$ ($8.5\%$ on average), while meeting performance demands. 


\end{abstract}
\section{Introduction}
\label{sec:intro}

\begin{figure}[!h]
\begin{center}
\includegraphics[trim=.5cm .5cm .7cm .5cm, clip=true,width=1\linewidth,keepaspectratio]{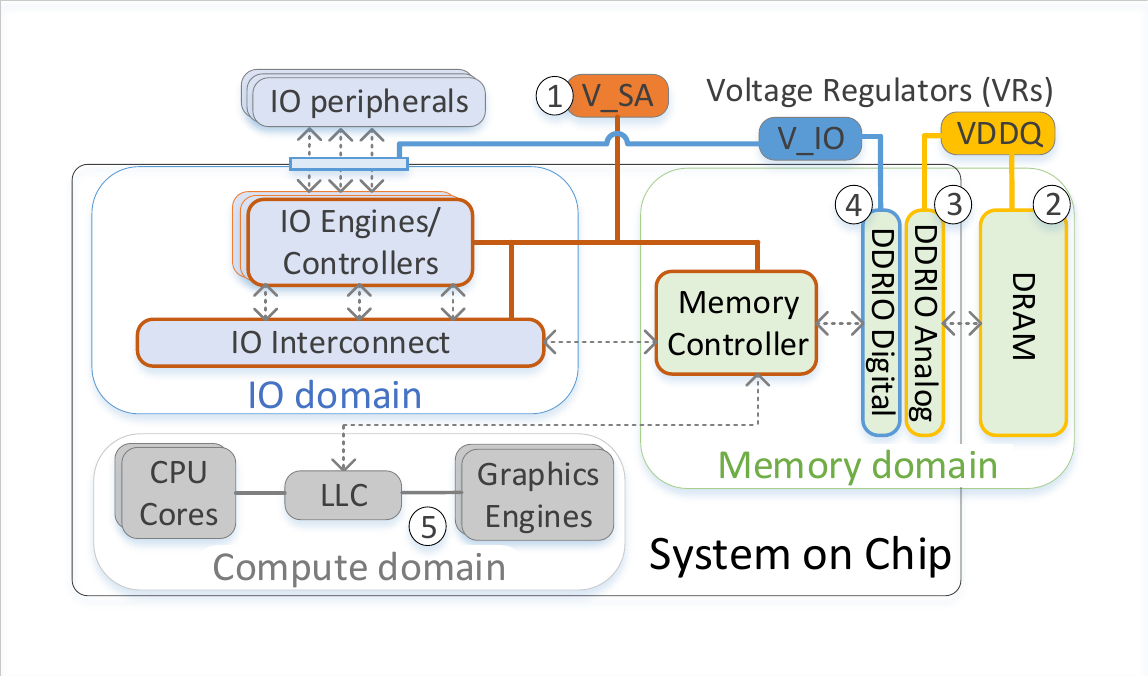}\\
\caption{\agy{A modern mobile SoC (}Intel Skylake \cite{21_doweck2017inside}) \agy{with three domains (}compute, IO, memory\agy{).} Voltage regulators (VRs) are highlighted, e.g., IO engines/controllers, \hjk{IO} interconnect, and memory controller share the same VR, V\_SA.}\label{skl_arch}
\end{center}
\end{figure}

A high-end mobile microprocessor is built as a system-on-chip (SoC) that integrates multiple components into a single chip. 
It typically has \emph{three} main domains: \emph{compute} (e.g., CPU cores, graphics engines), \emph{IO} (e.g., display controller, image signal processing (ISP) engine), and \emph{memory} (i.e., memory controller, memory interface, and DRAM) as illustrated in Fig. \ref{skl_arch}. 
A mobile SoC operates in a thermally-constrained environment, limited by what is known as thermal design power (TDP) \cite{gronowski1998alpha,TDP,rotem2009multiple,rotem2012power,rotem2013power,rotem2015intel,monchiero2006design,donald2006techniques,li2006cmp,skadron2004temperature,huang2008many,naveh2006power,esmaeilzadeh2011dark}.
To keep the system running below a TDP, the SoC power-management-unit (PMU) employs a \emph{power budget management} algorithm (PBM) to dynamically distribute the \agy{total} power budget to each SoC domain \cite{lempel20112nd,rotem2013power,rotem2012power,21_doweck2017inside,ranganathan2006ensemble,9_rotem2011power,nathuji2009vpm, bhojwani2007sapp, zhang2016maximizing,isci2006analysis}. This allows each domain to operate within its allocated power budget.
For instance, CPU cores and graphics engines in the compute domain share the same power budget. When a graphics-intensive workload is executed, the graphics engines consume most of the compute domain's power budget. 
To keep the power consumption of the compute domain within its allocated power budget, PMU applies  dynamic voltage and frequency scaling (DVFS) to 1) reduce the CPU cores' power consumption and 2) increase \hjk{the} graphics engines' \hjk{performance}   \cite{rotem2013power,rotem2012power,efraim2014energy,wu2005voltage,kim2008system,mallik2006user,wang2009predictive,komoda2013power,pathania2014integrated,rotem2015intel}.

\vspace{-1pt}
In this work, we demonstrate that the power budget  the PBM allocates to the IO and memory domains is \emph{inefficiently} managed, making the energy and performance of a high-end mobile SoC suboptimal. We make \agy{four} key observations\agy{.}

\noindent \textbf{Observation 1.}
In a typical high-end mobile SoC, the power budget management algorithm assigns a \emph{fixed} power budget to the IO and memory domains corresponding to the worst-case performance demands (bandwidth/latency) from the \emph{IO interconnect} and the memory subsystem resources. However, we observe that common use cases of mobile systems have only modest demands relative to the worst-case. 
Unfortunately, these systems do not apply DVFS to the IO and memory domains based on the \emph{actual} demands of the three domains, making these SoCs energy inefficient. 

\noindent \textbf{Observation 2.}
While mobile SoCs employ a power budget redistribution mechanism between components within a domain, such as between cores and graphics engines in the compute domain \cite{rotem2013power,rotem2012power,efraim2014energy}, we observe that current power budget management algorithms \emph{do not} support dynamic power redistribution \emph{across} different domains. 
Therefore, when a domain's power budget is underutilized, the remaining budget is wasted, making system performance suboptimal.
This unused power budget could have been allocated to another domain (e.g., the compute domain) to increase performance. 

\noindent \agy{\textbf{Observation 3.} In modern mobile SoCs, we observe that multiple components in the IO 
and compute 
domains have widely-varying main memory bandwidth demands across different workloads. However, due to over-provisioning \hjk{of} IO and memory demands, SoC energy efficiency remains low while running many workloads\hjk{,} as we demonstrate in Sec.~\ref{sec:motiv}.}

\noindent \agy{\textbf{Observation 4.} Unoptimized DRAM configuration register values can significantly reduce the energy efficiency benefits of multi-domain DVFS (e.g.\hjk{, they provide} $22\%$ less power reduction than optimized values).}

Unfortunately, there are \emph{three} main challenges that make it difficult for existing high-end mobile systems to apply DVFS to the IO and memory domains based on demands of  \emph{multiple} domains.
First, accurate prediction of  
1) the actual bandwidth/latency demands of the multiple domains, and 
2) the potential effect of DVFS on power/performance of the SoC, in the presence of multiple domains, is challenging. A modern high-end SoC integrates several components that share the IO interconnect and memory subsystem. Some of these components have strict quality of service (QoS) requirements \cite{usui2016dash}
with respect to latency (e.g., isochronous traffic \cite{benini2006networks,weber2005quality,dally2007interconnect}) and bandwidth (e.g., display\cite{intel_display,castellano2012handbook,usui2016dash,song2019self}). Mispredicting a component's actual demand can violate the QoS requirements and/or significantly degrade system performance.
Second, the DVFS process of the IO and memory domains is a global system optimization. It requires monitoring the demands of the three SoC domains and subsequently configuring multiple components in the SoC to carry out the actual DVFS. Therefore, a power management transition flow for applying this global optimization can be 
computationally expensive. 
If it is not done correctly, the transition from one voltage/frequency operating point to another can degrade SoC performance by stalling the SoC domains.
Third, the DVFS process should be holistic, efficient, and optimized to maximize power savings. 
For instance, previous works on memory subsystem DVFS \cite{david2011memory,deng2011memscale,chen2011predictive,deng2012coscale,felter2005performance,li2007cross,zhang2016maximizing,chang2017understanding,imes2018handing} do not \hjg{dynamically}  optimize \hjk{the} DRAM interface (i.e., DDRIO) \emph{configuration registers} \cite{mrc1,mrc2,mrc3} and voltage during the DVFS process. Unoptimized DRAM configuration registers and voltage can significantly reduce, or even negate, the potential power/performance benefits 
of memory subsystem DVFS, as we show in this paper (Sec. \ref{sec:motiv}). 


Recent works in \hjg{memory} DVFS 
\cite{david2011memory,deng2011memscale,chen2011predictive,deng2012coscale,felter2005performance,li2007cross,zhang2016maximizing,imes2018handing,deng2012multiscale} for modern 
SoCs focus only on  improving energy efficiency of
a \emph{single} domain (or limited components of two domains) and  do not address all three challenges mentioned above.
For example, MemDVFS~\cite{david2011memory} and MemScale~\cite{deng2011memscale} focus only on improving the energy efficiency of the main memory subsystem. CoScale~\cite{deng2012coscale} and other works~\cite{chen2011predictive,felter2005performance,li2007cross} consider coordinating the DVFS
of \agy{only} CPU cores and the main memory subsystem. 
To our knowledge, no previous work on SoC DVFS \agy{1) coordinates and} combines DVFS across three domains, or 2) optimizes the DRAM configuration registers \cite{mrc1,mrc2,mrc3} and voltage.

To enable more holistic power management in a mobile SoC and thereby to improve overall \hjk{SoC} power efficiency and performance,
we propose \emph{\tech}, a new power management \agy{technique}. 
{\tech} is based on three \textbf{key ideas}.
First, {\tech} can accurately and dynamically predict the bandwidth/latency demand of multiple SoC domains by implementing new performance counters and utilizing existing system configuration registers.
Second, {\tech} uses a highly\agy{-}efficient global DVFS mechanism to dynamically distribute the SoC power budget across all three domains, according to the predicted performance requirements. {\tech}'s DVFS mechanism minimizes latency overheads by 1) performing DVFS simultaneously in all domains to overlap the DVFS latencies and 2) storing the required configuration registers in on-chip SRAM near each domain.
Third, to maximize power savings, {\tech} optimizes the energy efficiency of \emph{each} domain at different DVFS operating points with domain-specific mechanisms. For instance, we optimize the energy efficiency of the DRAM interface by adding a dedicated scalable voltage supply and optimiz\agy{ing the} configuration registers for each DVFS operating point.

This work makes the following \textbf{contributions}:
\begin{itemize}

\item To our knowledge, {\tech} is the first work to enable coordinated and highly\agy{-}efficient DVFS across all SoC domains to increase the energy efficiency of mobile SoCs.
{\tech} introduces the ability to 
redistribute the \agy{total} power
budget across all SoC domains according to the performance demands of each domain.



\item We propose an effective algorithm to accurately predict the performance (e.g., bandwidth and latency) demands of the three SoC domains\agy{,} utilizing newly-implemented performance counters and  existing system configuration registers.
 
\item We introduce a new global DVFS mechanism that minimizes the performance overhead of applying DVFS across multiple domains.

\item We implement {\tech} on \hjk{the} Intel Skylake \hjg{SoC} for mobile devices \cite{tam2018skylake,anati2016inside,21_doweck2017inside} and evaluate {\tech} using a wide variety of \agy{workloads:} SPEC CPU2006 \cite{SPEC}, graphics (3DMark \cite{17_3DMARK}), and battery life workloads \agy{for} mobile devices~\cite{19_MSFT} (e.g., web browsing, light gaming, video conferencing, and video playback). On a 2-core Skylake with a $4.5W$ TDP, {\tech} improves the performance of SPEC CPU2006 and 3DMark workloads by up to $16\%$ and $8.9\%$ ($9.2\%$ and $7.9\%$ on average), respectively. 
For battery life workloads, which typically have fixed performance demands, {\tech} reduces the average power consumption by up to $10.7\%$ ($8.5\%$ on average), while meeting performance demands.
As the TDP reduces,  {\tech}'s relative benefits significantly increase. \hjg{For example, for a $3.5W$ TDP system, {\tech} enables \hjk{up to $33\%$ (}$19\%$ \hjk{on} average\hjk{)}  performance improvement o\hjk{n} SPEC CPU2006 workloads}.


\end{itemize}

\section{Background}
\label{sec:bg}

We provide a brief overview of \nv{a} modern mobile \hjg{SoC} architecture and
\agy{memory subsystem\hjk{,} with a focus \nv{on} power consumption and DVFS.}

\subsection{Mobile \hjg{SoC} Architecture} 
\noindent \textbf{Main Domains.} \agy{A high-end} mobile processor is \agy{a system-on-chip (SoC) that typically integrates three main domains into a single chip, as Fig.~\ref{skl_arch} shows\hjk{:}}
\agy{1)} compute (e.g., CPU cores and graphics engines), \agy{2)} IO (e.g., display controller, \hj{ISP engine}, IO interconnect), and \agy{3)} memory (i.e., \nv{the} memory controller, DRAM interface (DDRIO)\hjk{, and DRAM})\agy{.}
\agy{\nv{The} CPU cores, graphics engines, and IO controllers share the memory subsystem. Similarly, \nv{the} IO controllers share the IO interconnect.}

\noindent \textbf{Clocks and Voltages.} In \hjk{a modern} high-end mobile \hjg{SoC}, each one of the IO controllers/engines, IO interconnect, memory controller, and DDRIO typically have an independent clock. However, in current systems, there are three main \agy{voltage} sources for the IO and memory domains.  
First, the IO controller, IO interconnect, and memory controller \emph{share} the same voltage regulator, denoted as V\_SA\footnote{SA stands for System Agent which houses the traditional Northbridge \hjg{chip}. \hjg{SA} contains several functionalities, such as the memory controller and the IO controllers/engines~\cite{sunrise_point_skl}.}~\hjk{\circled{1}} in Fig. \ref{skl_arch}. 
Second, DRAM~\hjk{\circled{2}} and \agy{the analog part of the DRAM interface} \agy{(}DDRIO-analog\agy{)}~\hjk{\circled{3}} share the same voltage \agy{regulator}, known as VDDQ. 
Third, the digital part of the DRAM interface (DDRIO-digital)~\hjk{\circled{4}} typically shares the same voltage \nv{as} the IO interfaces (\agy{e.g.,} display IO, ISP IO), denoted as V\_IO in Fig. \ref{skl_arch}.     
%
The \nv{c}ompute domain\hjk{~\circled{5}} typically has two voltage source\nv{s} (not \nv{depicted} in Fig. \ref{skl_arch}): 1) a voltage regulator that is shared between CPU cores and LLC and 2) a voltage regulator for the graphics engines \cite{singh20173,singh2018zen,burd2019zeppelin,toprak20145,2_burton2014fivr,5_nalamalpu2015broadwell,tam2018skylake,qcom2018,nikolskiy2016efficiency}. 

 \subsection{Memory System Organization}

We provide a general overview of the structure of \agy{a} DRAM device.
A set of DRAM chips placed together on a \agy{DRAM module} comprises a rank. A DRAM chip is typically divided into multiple banks along with IO hardware (drivers and receivers) \nv{that} enables access to the contents of the storage cells from outside the chip via the memory interface (DDRIO). Each bank includes peripheral logic to process commands and a grid of rows (wordlines) and columns (bitlines) of \agy{DRAM} cells. \agy{A DRAM cell consists of a capacitor and an access transistor. A single bit of data is \hjk{encoded by the} charge level \hjk{of} the capacitor.} \agy{For more detail, we refer the reader to prior works on DRAM organization and operation \hjk{(e.g.,} \cite{kim2012salp,lee2015simultaneous,lee2015decoupled,lee2013tiered,seshadri2013rowclone,hassan2019crow,seshadri2019dram,raidr,mutlu2007stall,mutlu2009parallelism,hassan2016chargecache, seshadri2017ambit,ghose2018your,ghose2019demystifying,lee2015adaptive,hassan2017softmc}).}

\subsection{Memory Power Consumption}

The power consumption of a DRAM system comprises background power, operation power, and \agy{the memory controller} power, \agy{which we briefly describe}. We refer the reader to \cite{david2011memory, ghose2018your, chang2017understanding, ghose2019demystifying} for a more detailed overview. 

\noindent \textbf{Background Power.} A DRAM chip continuously consumes a certain amount of power in the background with or without memory accesses. The background power consumption has two main \nv{sources}. First, the maintenance task that the peripheral circuitry performs to ensure signal integrity between the processor and DRAM chip. Second, periodic refresh operations that restore a leaky DRAM cell's charge level to ensure that a cell does not leak enough charge to cause a bit flip~\cite{ghose2018your, raidr,liu2013experimental,patel2017reach}.


\noindent \textbf{Operation Power.}
\agy{Operation power is a DRAM chip's active power consumption when it executes memory commands for a memory access.}
\hjk{It} includes the DRAM array power, IO power, register power, and termination power.

DRAM array power is consumed by the core of the memory (e.g., banks, row\agy{/}column decoders, and sense amplifiers). Thus, \agy{DRAM} array power consumption correlates with \agy{DRAM} access count (\agy{i.e.,} memory bandwidth \agy{consumption}). The array draws a constant active-state power when a read, write\nv{,} or precharge command is active.
The IO power is consumed by input buffers, read/write latches, delay-locked loop (DLL), data interface drivers, and control logic that are used to transfer data from/to a DRAM chip. The register power is consumed by the \hjg{input/output} registers \hjg{placed} on clock and command/address DRAM interface lines, 
and their \hjg{clock circuits (e.g., phase-locked loop (PLL))}. 
The termination power includes the power used in terminating the DDRIO during active operation. Termination power depends on interface utilization and it is not directly frequency-dependent~\cite{david2011memory}.


\noindent \textbf{Memory Controller Power.} \agy{M}emory controller power is the combination of \agy{1)} the static power consumption, \agy{which is} proportional to operational $voltage$ and $temperature$, and \agy{2)} the dynamic power consumption, \agy{which is proportional to} $voltage^2 \times frequency$ of the memory controller. 

\subsection{Memory DVFS}
Dynamic voltage and frequency scaling (DVFS) is a technique that reduces the power consumption of an SoC component (e.g., a CPU core, a graphics engine) by reducing its voltage and frequency\nv{,} \agy{potentially} at the expense of performance \cite{haj2018power,gough2015cpu,david2011memory,deng2011memscale,chen2011predictive,deng2012coscale,felter2005performance,li2007cross,zhang2016maximizing,imes2018handing,deng2012multiscale}.
While DVFS can allow \agy{a} quadratic reduction in energy consumption and a cubic reduction in average power dissipation at the expense of a linear reduction in performance \nv{of} the SoC component, the overall system energy consumption may \emph{increase} due to longer execution time and higher utilization (i.e., less time spent in the idle power state) of other system components \cite{haj2018power,gough2015cpu}. Therefore, techniques such as race-to-sleep \cite{zhang2017race,efraim2014energy} increase the frequency (and voltage) of an SoC component to reduce system\agy{-}level energy. System architects use various metrics, such as the energy delay product (EDP\footnote{Energy-delay product (EDP) \cite{gonzalez1996energy} is a commonly used metric for quantifying a \hjn{computing} system's energy efficiency. The lower the EDP the better the energy efficiency.}) to measure energy efficiency \cite{gonzalez1996energy,haj2018power}, in a way that combines both energy and performance (delay).



The main objective of using DVFS in the memory subsystem of a mobile \hjg{SoC} is to improve energy efficiency (i.e., reduce EDP). To do so, memory DVFS reduces the background, operation, and \agy{memory controller} power consumption by reducing the frequency and voltage of the memory subsystem.
Current approaches scale the frequencies of the \agy{memory controller}, DDRIO \hjk{(both analog and digital)}, and DRAM device, while the voltage is reduced only for the \agy{memory controller}.
Modifying the operating voltage of the \agy{DDRIO\hjk{-analog}} and \hjk{the} DRAM
\agy{device} is \emph{not} yet supported by commercially-available\footnote{There are substantial challenges in operating a DRAM array at multiple voltages, which requires precisely tuned timing, transistor sizing, and voltage partitioning \cite{JEDEC_dvfs}.} DRAM devices \cite{chang2017understanding,deng2011memscale,JEDEC_dvfs}\agy{.} \agy{T}o maximize the energy savings while applying memory DVFS, we also \hj{concurrently} apply DVFS to \agy{DDRIO-digital (\circled{4} in Fig.~\ref{skl_arch})} \hjg{and} the IO \agy{interconnect}. 

\noindent \textbf{Impact of Memory DVFS on the SoC.} 
Reducing the frequency of the memory subsystem affects \hjk{the} power consumption and performance of the SoC. \emph{Performance} is affected because reducing the frequency 1) makes data bursts longer, 2) increase\nv{s} memory access time \nv{as it slows down both the} memory controller and \hjk{the} DRAM interface, and 3) increase\nv{s} the queuing delays at the memory controller.

\emph{Power and energy consumption} \hjk{are} also affected by reducing the memory subsystem frequency in four \nv{ways}. 
First, background power reduces linearly. 
Second, memory controller power reduce\hjk{s} approximately by a cubic factor due to the reduction of memory controller voltage (V\_SA \circled{1} in Fig. \ref{skl_arch}) \cite{haj2018power,gough2015cpu,david2011memory,deng2011memscale,chen2011predictive,deng2012coscale,felter2005performance,li2007cross,zhang2016maximizing,imes2018handing,deng2012multiscale}. 
Third, lowering the DRAM \hjg{operating} frequency increases read, write\nv{,} and termination energy linearly, because each access takes longer time. Fourth, due to the degradation in performance, the utilization of other components in the system can increase, which increases the overall SoC energy.

\subsection{Memory Reference Code}\label{sec:mrc}
Memory reference code (MRC \cite{mrc1,mrc2,mrc3}) is part of the BIOS code that manages system memory initialization. The purpose of  MRC training is to 1) detect the DIMMs and their capabilities, 2) configur\nv{e} the configuration registers (CRs) of \nv{the} memory controller (MC), DDRIO\nv{,} and DIMMs, and 3) train the data and command interfaces (DDRIO) for correct operation and optimized performance \hj{as defined by \nv{the} JEDEC standard \cite{JEDEC_training}}. 

MRC \hjk{training} is typically carried out with a \emph{single} DRAM frequency. Therefore, the configuration register values of MC, DDRIO, and DIMMs are optimized for this particular DRAM frequency. 
When dynamically switching the memory subsystem between multiple frequencies \hj{(e.g., during DVFS)}, \hj{the configuration register values} should be updated \hj{to the optimized \agy{values} corresponding to the new DRAM frequency}. Otherwise\nv{,} the \hjk{unoptimized} \hj{configuration registers} \hj{\agy{can} degrade performance and negate
potential \agy{benefits} of DVFS\nv{,} as we show in Sec. \ref{sec:motiv}}.  
\section{Motivation}
\label{sec:motiv}
To experimentally motivate building {\tech}, we carry out an experiment on the Intel Broadwell processor \cite{5_nalamalpu2015broadwell}, the previous generation of our target Skylake processor \cite{21_doweck2017inside}.  
The goal of this experiment is to \nv{evaluate} the potential benefits \nv{of employing} DVFS across three 
SoC domains.
\hjg{We use multiple workloads from SPEC CPU2006 \cite{SPEC} and 3DMark \cite{17_3DMARK}, and a workload that exercises the peak memory bandwidth of DRAM \cite{mccalpin1995memory}.}
We use two setups for our experiment, as we show in Table \ref{tab:motiv_two_cfg}:
1) To examine the performance of the processor without applying DVFS across multiple domains,
we use a \emph{baseline setup} in which we set the CPU core frequency to $1.2GHz$ and maintain the default voltage and frequency values of the other SoC components (e.g., DRAM, IO interconnect, and DDRIO digital). 
2) To \nv{evaluate} the benefits of applying DVFS across multiple domains, we use a \emph{multi-domain DVFS setup} (MD-DVFS) in which the CPU cores have the same frequency as the first setup, but we reduce the frequency and voltage values of other SoC components in the IO and memory domains.

\begin{table}[!h]
\centering
\caption{Our two real experimental setups 
}
\label{tab:motiv_two_cfg}
\begin{tabular}{lcc}
\hline
\textbf{Component}       & \textbf{Baseline} & \textbf{MD-DVFS} \\ \hline
DRAM frequency                  & 1.6GHz            & 1.06GHz                \\
IO Interconnect & 0.8GHz            & 0.4GHz                 \\
Shared Voltage       & V\_SA             & $0.8\cdot$V\_SA              \\
DDRIO Digital        & V\_IO             & $0.85\cdot$V\_IO              \\
2 Cores (4 threads)  & 1.2GHz            & 1.2GHz                 \\ \hline
\end{tabular}
\end{table}

We attain the reduced-performance setup by changing \emph{four} parameters. 
First, we reduce the DDR frequency by one bin\footnote{DRAM devices support a few discrete frequency bins (normally only three). For example, \hjk{LP}DDR3~\cite{no2013jesd79} supports only $1.6GHz$, $1.06GHz$, and $0.8GHz$. The default bin for most systems is the highest frequency \cite{5_nalamalpu2015broadwell,21_doweck2017inside}.} (i.e., $1.06GHz$). Doing so proportionally reduces the memory controller (MC) frequency that normally operates at half the DDR frequency. Since MC and IO interconnect share the same voltage (V\_SA \circled{1} in Fig. \ref{skl_arch}), we also proportionally reduce the IO interconnect clock frequency to align it with the voltage levels 
of the IO and memory domains \hj{based on the voltage/frequency curves of both domains}. 
Second, we reduce the shared voltage (V\_SA \circled{1}) and DDRIO-digital voltage (V\_IO~\circled{4}) proportionally to the minimum functional voltage corresponding to the new frequencies in the IO and memory domains. 
Third, we maintain the CPU core~\circled{5} frequency and VDDQ~\circled{3} voltage unchanged across the two experimental setups.
Fourth, to optimize the power/performance of the SoC in each setup, we configure the DRAM device, MC, and DDRIO with optimized MRC values for  the selected DRAM frequency, as we explain in Sec. \ref{sec:mrc}. 
Based on our evaluation of the SoC using the two setups, we make the following four key observations.

\noindent \textbf{Observation 1.}
Current mobile systems have substantial 
\hj{energy} inefficiency with respect to managing \hj{the voltage and frequency} of the IO and memory domains.
 
\begin{figure}[!h]
  \begin{center}
  \includegraphics[trim=.7cm 0.7cm .9cm .5cm, clip=true,width=1.\linewidth,,keepaspectratio]{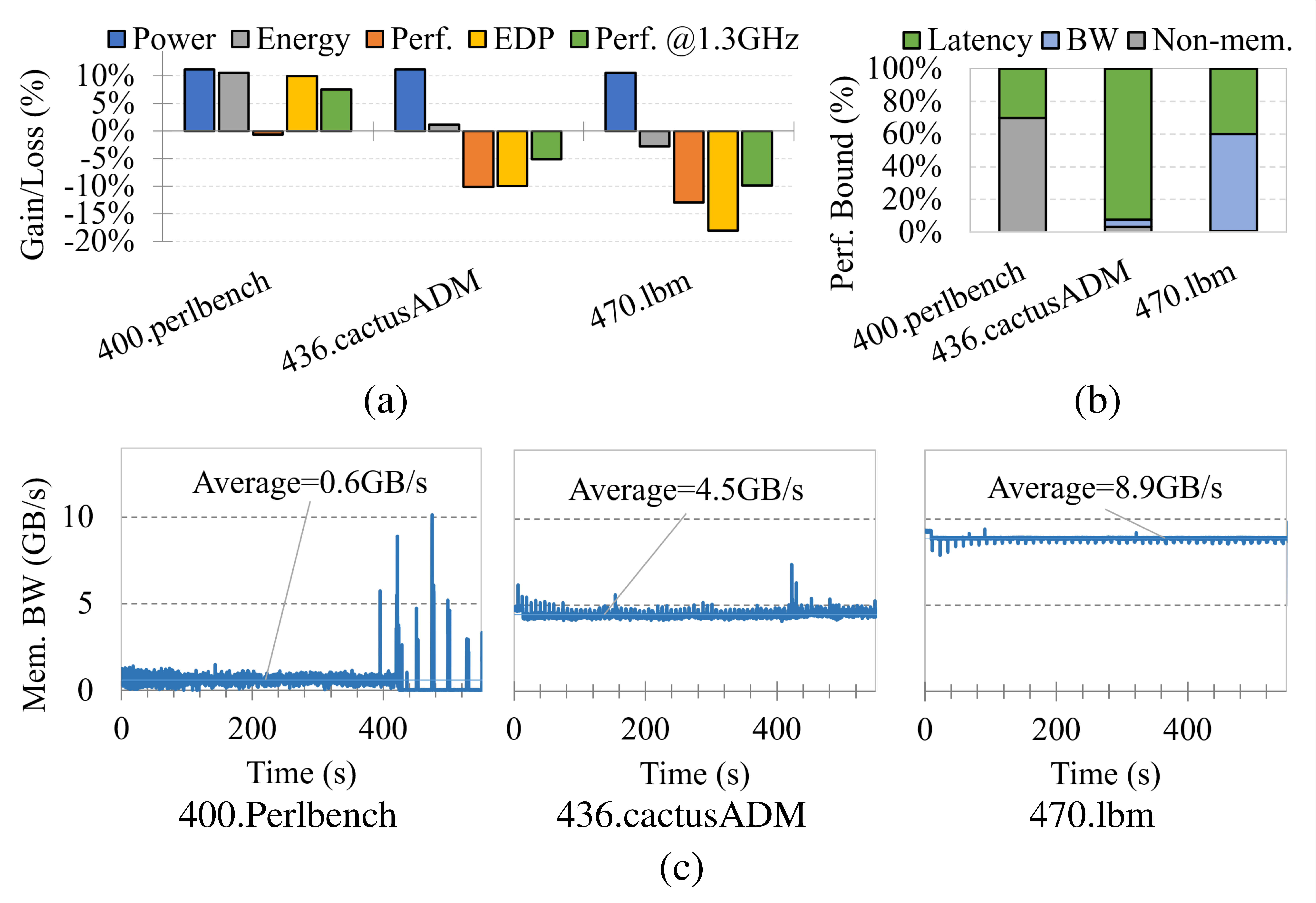}\\
  \caption{(a) \ch{Summary of the impact of MD-DVFS over the baseline on different metrics.} (b) Bottleneck analysis of the three workloads\hjk{, showing} what fraction of the performance is bound by main memory latency, main memory bandwidth or non-main memory related events. (c) Memory bandwidth (BW) demand of the three benchmarks. 
  }\label{fig:avgp_perf_energy}
  \end{center}
 \end{figure}

We show in  Fig. \ref{fig:avgp_perf_energy}(a) the impact of the MD-DVFS setup on the average power consumption, energy consumption, performance, and energy-delay-product (EDP) compared to the baseline (without applying \hjn{DVFS}) using the three SPEC CPU2006 benchmarks. Fig. \ref{fig:avgp_perf_energy}(a) also shows the performance \hjk{of the MD-DVFS setup}, compared to the baseline setup, when we increase the CPU core frequency from $1.2GHz$ to $1.3GHz$.

\ch{We make three key observations from Fig.~\ref{fig:avgp_perf_energy}(a): First, the average power consumption of \emph{all} three benchmarks reduce\hjk{s} (by $10\%$--$11\%$) with MD-DVFS. Second, while power consumption reduces in all evaluated workloads, in several workloads (e.g., cactusADM and lbm), there is a significant loss in performance (>$10\%$) as result of the reduction in frequency of the memory domain. Third, the effect of energy consumption varies widely across workloads. Workloads such as \hjg{perlbench}
\hjg{has reduced energy consumption by \hjg{$11\%$}.}  
For workloads such as \hjg{cactusADM and lbm,}
which experience a performance degradation \hjk{with} MD-DVFS, energy consumption only improves slightly or increases. 
}

\ch{Fig.~\ref{fig:avgp_perf_energy}(b) shows a bottleneck analysis of the same three workloads evaluated in Fig.~\ref{fig:avgp_perf_energy}(a).
We plot} what fraction of the performance is bound by main memory latency, main memory bandwidth or non-main memory related events.
We observe from Fig. \ref{fig:avgp_perf_energy}(b) that 
 cactusADM and lbm are heavily main memory bottlenecked, and the main bottleneck in cactusADM is main memory latency.
Fig.~\ref{fig:avgp_perf_energy}(c) plots the memory bandwidth demand of the three benchmarks, showing that the memory bandwidth demand varies both over time and across workloads. 
Overall, the core-bound perlbench has low demand, but demand spikes at times, lbm has constant high demand, and cactusADM has moderate demand (but even then MD-DVFS hurts its performance by more than $10\%$).


\ch{We conclude that for workloads that are not memory latency or bandwidth bound, scaling down the voltage and frequency of the memory and IO domains can significantly reduce power and energy consumption with minimal effect on performance. }

\noindent \textbf{Observation 2.} 
While mobile SoCs employ a power budget redistribution mechanism between \hj{SoC} components within a domain, such as between CPU cores and the graphics engine\hjk{s} in the compute domain \cite{rotem2013power,rotem2012power,efraim2014energy}, we \ch{observe} that current mobile SoCs \emph{do not} support dynamic power redistribution across different domains. Fig.~\ref{fig:avgp_perf_energy}(a) shows the performance impact of increasing the CPU cores' frequency of the MD-DVFS setup from $1.2GHz$ to $1.3GHz$ when reassigning the saved average power budget from \hj{the IO and memory domains to the compute domain}.
Performance of perlbench improves significantly by $8\%$ over the baseline with  MD-DVFS. \ch{Workloads that are not limited by compute bandwidth \hjg{(e.g., cactusADM and lbm)} 
do not \hjk{experience} performance improvement \hjk{over the baseline} \hjm{with higher} core frequency. }

We conclude that 1) scaling down the voltage and frequency of the IO and memory domains when demands from these domains is low and 2) redistributing the saved power budget between domains can 
improve performance \ch{in workloads that are compute bound}.

\begin{figure}[!h]
  \begin{center}
  \includegraphics[trim=.5cm .7cm .5cm .7cm, clip=true,width=1.0\linewidth,,keepaspectratio]{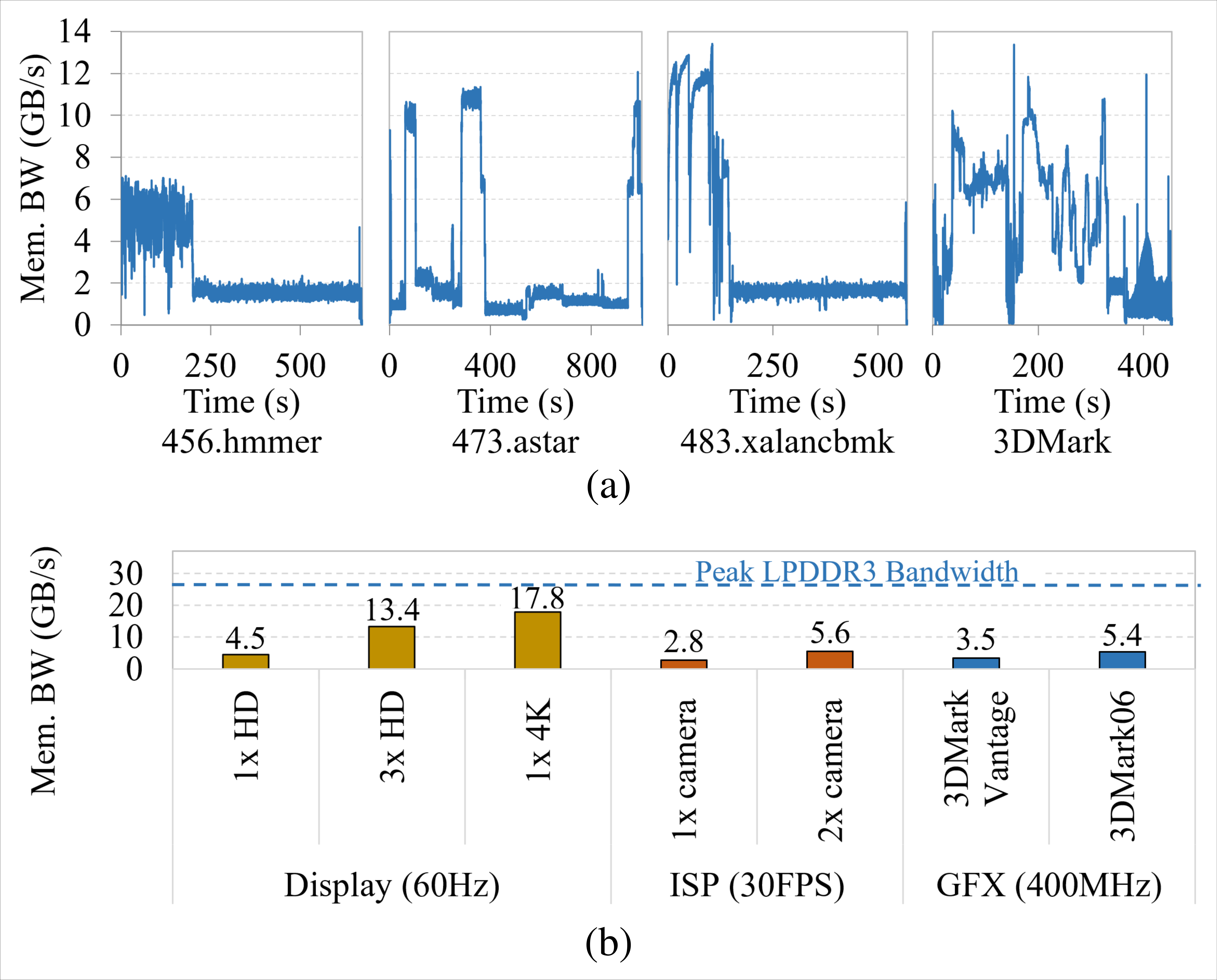}\\
  \caption{ 
  (a) Memory bandwidth (BW) demand over time for three  SPEC CPU2006 benchmarks and the 3DMARK\cite{17_3DMARK} graphics benchmark. 
  (b) Average memory bandwidth demand of \ch{the} display engine, ISP engine, and graphics engine\hjk{s} (GFX) using different configurations and workloads. 
  }\label{fig:mem_bw}
  \end{center}
 \end{figure}

\noindent \textbf{Observation 3.}
In modern mobile SoCs, we observe that multiple components in the IO \hj{(e.g., display controller, ISP engine)} and compute \hj{(e.g., CPU cores, graphics engine\hjk{s})} domains have widely-varying main memory bandwidth demands.
Fig. \ref{fig:mem_bw}\hj{(a)} illustrates this by showing how the main memory bandwidth demand of three  SPEC CPU2006 workloads and a 3DMark graphics workload varies over time. These workloads typically require different memory bandwidth over time.\footnote{The memory bandwidth demand was measured on the Intel Broadwell system \cite{5_nalamalpu2015broadwell} \emph{without} scaling the voltage or frequency of any domain or component.}
Fig. \ref{fig:mem_bw}\hj{(b)} shows the memory bandwidth demand of the display engine, ISP engine, and graphics engine\hjk{s} (GFX) using different configurations and workloads.
\hjg{We observe that the display engine has widely-varying memory bandwidth demands depending on the display quality: HD display consumes approximately 17\% of the peak \hjg{memory}  bandwidth of a dual-channel LPDDR3 ($25.6GB/s$ at $1.6GHz$ DRAM frequency), while a single 4K display (the highest supported display quality in our system) consumes 70\% of the peak \hjg{memory} bandwidth.}

\hjg{We conclude that typical workloads have modest demands yet the SoC IO and memory demands are provisioned high, making existing mobile SoCs energy inefficient for typical workloads.}

\begin{figure}[h]
  \begin{center}
  \includegraphics[trim=1.1cm .8cm .8cm 1.6cm, clip=true,width=.7\linewidth,,keepaspectratio]{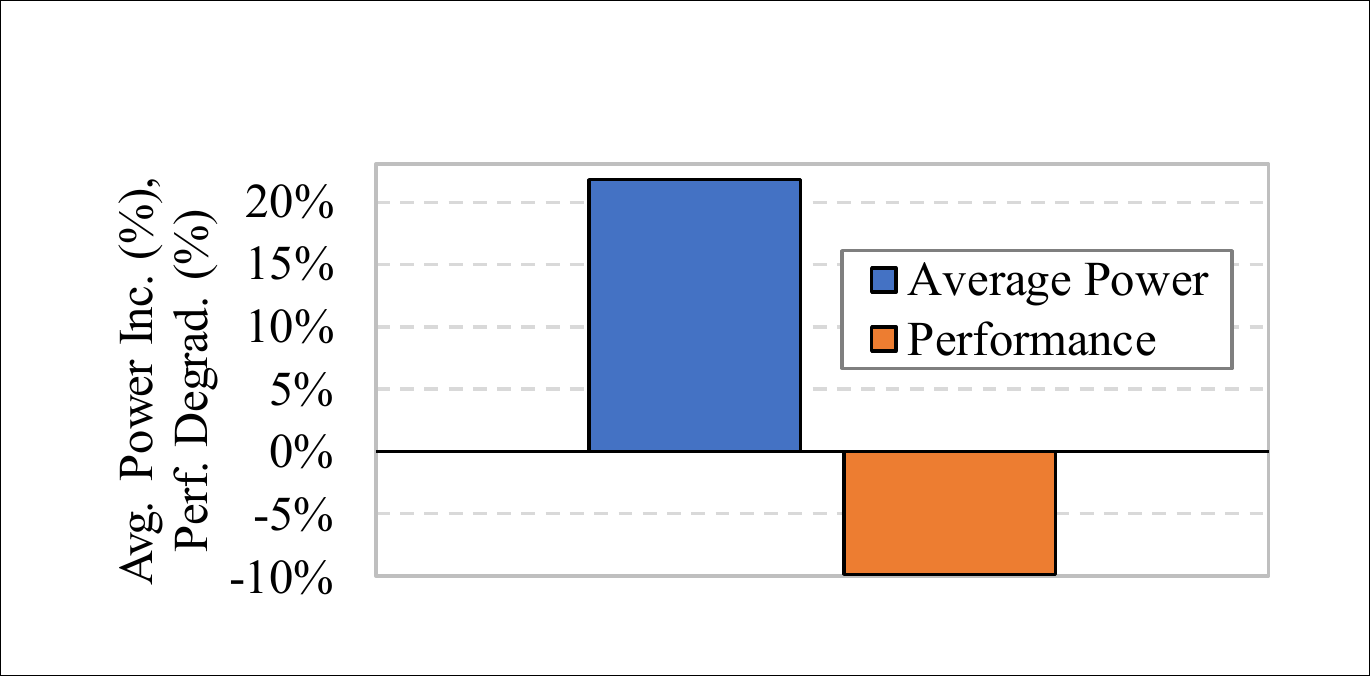}\\
  \caption{\ch{Impact of using unoptimized MRC values on power consumption and performance for a memory-bandwidth-intensive microbenchmark.}
  }\label{fig:opt_mrc}
  \end{center}
 \end{figure}

\noindent \textbf{Observation 4.}
We observe that choosing optimized MRC values for the DRAM configuration registers is \hj{important} for improving multi-domain DVFS energy efficiency. 
\ch{In Fig. \ref{fig:opt_mrc}, we show the impact of using \emph{unoptimized} MRC values on the overall performance and power consumption \hjk{of the MD-DVFS setup}. We use a microbenchmark that was designed to exercise the peak memory bandwidth of DRAM (similar to STREAM \cite{mccalpin1995memory,mutlu2007memory})}. 
\hjg{This helps us to isolate the impact of unoptimized MRC values on power/performance of the memory subsystem.} 
We observe that unoptimized MRC values can greatly degrade both average power (by $22\%$) and performance (by $10\%$) compared to using optimized MRC values.   
 
Based on our four key observations, we conclude that a \textit{holistic power management approach} is needed  to mitigate the \hjg{ power management} inefficiencies in current mobile SoCs. This holistic approach should 
1) redistribute the power budget \hjg{of SoC domains} based on the actual demands \hjg{of a workload from each domain}, 
2) simultaneously scale the frequency/voltage of all domains, and 
3) \ch{be optimized separately for each domain to minimize inefficiencies in the DVFS mechanism}
\hjg{(e.g., \hjg{by} using optimized MRC values \hjg{in the DRAM interface})}. 

\section{{\tech} Architecture}
\label{sec:technique}


We design {\tech} with two design goals in mind: 
1) reduce power consumption by dynamically scaling the voltage/frequency of all SoC domains \agy{based} on \agy{performance demands from different} domain\agy{s}, and
2) improve system throughput by redistributing the power budget across \agy{SoC} domains \agy{based on performance demands}. 

{\tech} achieves these two goals with \emph{three} key components. The \emph{first component} is a power management flow that is responsible for 1) scaling the multiple voltages and frequencies \agy{in} the \hjg{IO and memory} domains and 2) reconfiguring the DRAM interface with optimized MRC values for the selected DRAM frequency.

The \emph{second component} of {\tech} is a \emph{demand prediction mechanism} that uses both the system configuration and dedicated performance counters to predict the static and dynamic \agy{performance} demands \agy{from} the SoC domains. This component is important, as mispredicting the actual demand of the workload or \agy{an} IO device places the system in an improper DVFS operating point\agy{,} which can significantly degrade workload performance or violate quality of service requirements that the IO device might have. 

The \emph{third component} of {\tech} is a \emph{holistic power management algorithm} that is responsible for scaling the voltage and frequency of the IO interconnect and memory subsystem to meet the system's dynamic \hjk{performance} demand.
The power management algorithm \agy{uses} the additional power budget saved as a result of \agy{DVFS in the} IO and memory domains, to increase the performance of \agy{the compute domain} (\hjg{i.e., CPU cores and graphics engines}).
This leads to improved overall system throughput while maintaining the average system power within the thermal design power limit.

The three components \agy{of \tech{}} work together to orchestrate the voltage and frequency of each of the SoC domains and significantly reduce the overall energy consumption of a mobile SoC. \agy{W}e describe \agy{them} in detail \agy{in} the \agy{next} three \agy{subsections}. 



\subsection{Power Management Flow}
\begin{figure*}[h]
  \begin{center}
  \includegraphics[trim=.5cm 0.9cm .5cm .65cm, clip=true,width=1\linewidth,keepaspectratio]{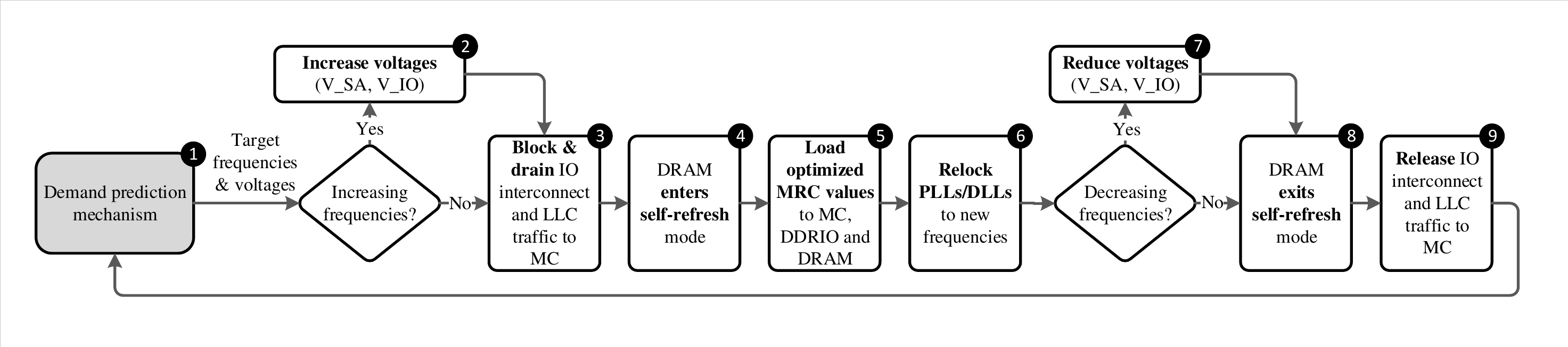}\\
  \caption{{\tech} power management flow  which carry out the DVFS of SoC domains. 
  }\label{fig:flow}
  \end{center}
  \vspace{-10pt}
 \end{figure*}

{\tech} power management flow\agy{, implemented by the PMU,} is responsible for adjusting the frequencies and voltages of the IO interconnect and memory subsystem\agy{, as depicted} in Fig. \ref{fig:flow}. First, {\tech}'s \emph{demand prediction mechanism} \circledb{1} (described \agy{in Sec.~\ref{subsec:demand_prediction_mechanism}}) \agy{initiates} a frequency change by determining new target frequencies/voltages for the SoC domains. If the demand prediction mechanism decide\agy{s} to increase (decrease) the IO and memory domain frequencies, then the flow increases (decreases)  the voltages before~\circledb{2} (after~\circledb{7}) the actual increase (decrease) of the PLL/DLL clock frequencies~\circledb{6}. Subsequently, the flow blocks and drains the IO interconnect and the traffic from cores/graphics into the memory domain (i.e., LLC traffic to memory controller) \circledb{3}. 
To safely block the interconnect, all the outstanding requests are completed and new requests are not allowed to \agy{use} the interconnect during this period. 
After all requests are completed, DRAM enters self-refresh mode~\circledb{4}, and the flow loads the new optimized MRC values for the new DRAM frequency from on-chip SRAM into the memory-controller, DDRIO and DRAM configuration registers~\circledb{5}. Next, \hjl{the} {\tech} flow sets the clocks to the new frequencies by re-locking both the \agy{p}hase-locked loop\hjk{s} (PLLs) and delay-locked loop\hjk{s} (DLLs) to the new IO interconnect and memory subsystem frequencies~\circledb{6}. 
Finally\hjl{,} DRAM \agy{exits} self-refresh mode~\circledb{8} and both the IO interconnect and the LLC traffic to the memory controller \circledb{9} \hjg{are released}. This resumes SoC execution with the new frequencies/voltages \agy{for SoC domains} and an \agy{o}ptimized DRAM interface for the new DRAM frequency.

\subsection{Demand Prediction Mechanism}
\label{subsec:demand_prediction_mechanism}
The {\tech} demand prediction mechanism uses peripheral configuration registers (e.g., the number of active displays or cameras) and new performance counters that we propose, to predict \hjk{the} performance demands of the SoC domains. We divide the performance demands \agy{in}to two categories: \emph{static} and \emph{dynamic} performance demands. We consider \agy{a} performance demand as static if it \agy{is only related} to system configuration (e.g., number of active displays or cameras). System configuration typically change\agy{s at the time-scale of tens of milliseconds \hjg{as it is normally controlled by software (e.g., OS, drivers)}.}
Therefore, the PMU has enough time to respond to any configuration change that requires, for example, transition from a low- to high-bandwidth operating point of {\tech}\agy{,} without affecting system performance or quality of service. Such a  transition can be completed within several \emph{microseconds}. 
Dynamic performance demands \hjg{are related to workload phase changes, which could happen much more frequently (e.g., \hjk{hundreds} of cycles)}.
Next, we explain how the \hjk{performance} demand prediction is performed based on each demand category.

\noindent \textbf{Static Performance Demand \agy{Estimation}}. 
To estimate static performance demand,  {\tech}  maintains a \emph{table} inside the firmware of the power-management unit (PMU) that maps every possible configuration of peripherals connected to the processor to \agy{IO and memory} bandwidth/latency demand values. 
The firmware obtains the current configuration from control and status registers (CSRs) of these peripherals. 
For example, modern laptops support up to three display panels \cite{msft_surface,macBook_air}. 
When only one display panel is active (connected), it requires a certain bandwidth, but when three of the same display panel are connected, the bandwidth demand is nearly three times higher\agy{,} as depicted in Fig. \ref{fig:mem_bw}(b). 
\hjg{In this case,} the number of active displays \hjl{and the} resolution and refresh rate for each display are available in the CSRs \agy{of} the display engine.

Our method enables accurate estimation of the static bandwidth \agy{demand} based \agy{solely} on the peripheral configuration \agy{as the bandwidth demand of a given peripheral configuration is known and \hjl{is} deterministic.}


 \begin{figure*}[t]
  \begin{center}
  \includegraphics[trim=.6cm .7cm .6cm 0.7cm, clip=true,width=1.0\linewidth,,keepaspectratio]{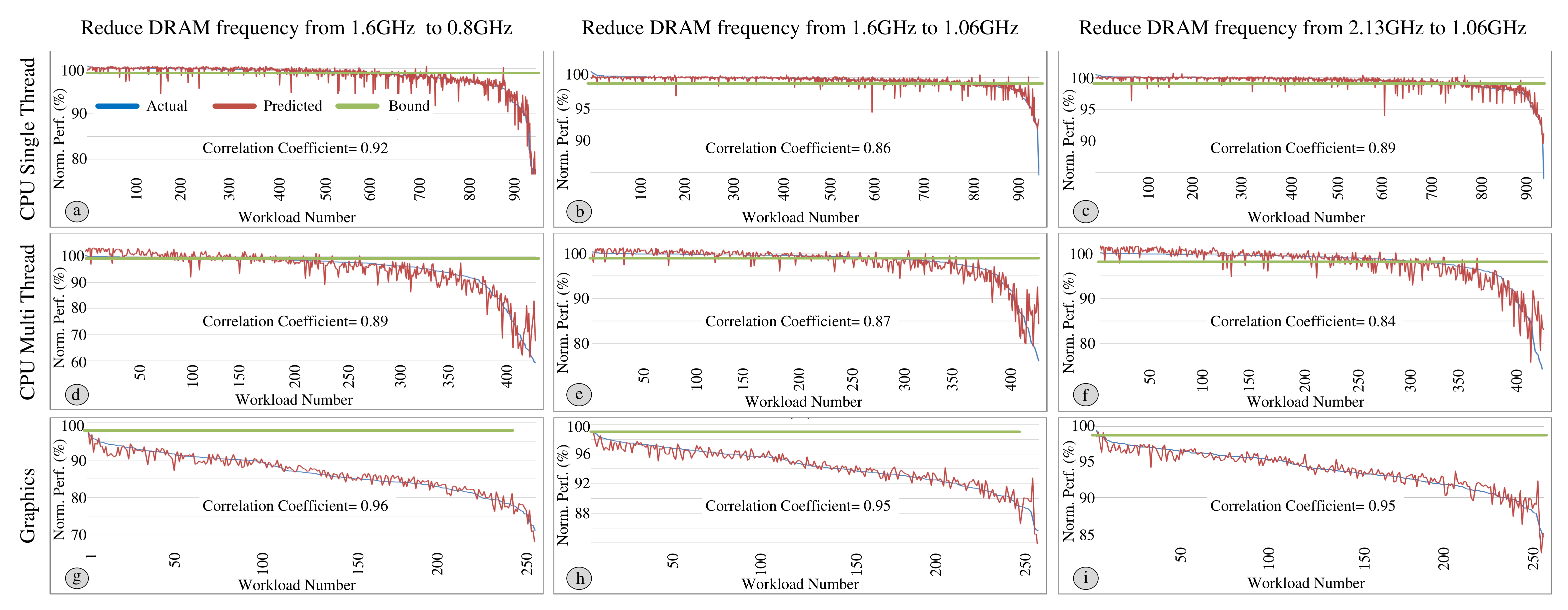}\\
  \caption{\hjm{Actual vs. Predicted performance impact of reducing} 
  \hjk{DRAM frequency} \hjm{in more than $1600$ workloads}. We examine three pairs (high/low) of DRAM frequencies: $1.6GHz$/$0.8GHz$, $1.6GHz$/$1.06GHz$ and $2.13GHz$/$1.06GHz$. 
  \hjm{Performance is normalized to that with the higher DRAM frequency.}
  \hjm{We also show the correlation coefficient between the Actual vs. Predicted performance impact values.}
  }\label{fig:predict_perf_impact}  
  \end{center}
  \vspace{-10pt}
 \end{figure*}

\noindent \textbf{Dynamic Performance Demand \hjg{Estimation}}. We categorize the bandwidth/latency demand \agy{of a} workload as dynamic as \agy{it changes over time.}
For instance, SPEC CPU2006 workloads \cite{SPEC} \agy{demand} different \agy{amounts of main} memory bandwidth over time\agy{,} as illustrated in Fig. \ref{fig:mem_bw}. 
\agy{Similarly, mobile workloads have dynamically varying memory latency demands, as shown in~\cite{haj2015doee}.}

We find that we are able to use \agy{existing} performance counters to predict the dynamic bandwidth \hjg{and latency} demand when running a workload at a reduced frequency.
We \emph{select performance counters} for predicting dynamic \agy{IO and \hjg{main} memory bandwidth \hjk{and} \hjg{latency}} demand\agy{s}
using two steps: 1) among tens of internal processor \hjk{performance} counters, we build an initial selection based on intuition, and 2) we empirically prune our selection using an iterative process until the correlation between the performance counters and the workload's performance degradation is closer to our target (e.g., ${>}90\%$ of our target). 
In the first stage, we choose an initial set of \hjl{($15$)} internal performance counters related to memory requests from the three domains of the SoC (i.e., \agy{c}ompute, IO, and memory). 
Subsequently, we run \hjm{a large number} of representative mobile \textit{\hjk{workload}s}\footnote{The  \hjk{workload}s comprise of representative performance and office productivity workloads including SPEC06\cite{SPEC}, SYSmark\cite{sysmark}, MobileMark\cite{mobilemark}, \agy{and} computer graphics intensive workloads (e.g., 3DMARK\cite{17_3DMARK}).} with the two setups, baseline and \agy{multi-domain} DVFS (MD-DVFS). We run the \hjk{workload}s  on a real system with \agy{the same} setup \agy{shown in} Table \ref{tab:motiv_two_cfg}\hjl{,} while selecting multiple DRAM frequencies for both \hjk{the} \agy{baseline} and the MD-DVFS setup.
We examine the performance of each run in addition to the values of the performance counters. Fig. \ref{fig:predict_perf_impact} shows the actual versus the predicted performance impact of reducing DRAM frequency from a baseline frequency to a lower frequency  used in MD-DVFS setup when running \hjm{more than $1600$} \hjk{workload}s. The figure also shows the \hjk{correlation} coefficient of the prediction when using the final list of performance counter\hjk{s}. 
The final list of performance counters used in \agy{our prediction algorithm} is as follows:
\begin{itemize}
    \item 	\textbf{GFX\_LLC\_MISSES} counts the number of Last-Level-Cache (LLC) misses due to \hjl{graphics engines} (GFX). This performance counter is indicative of the bandwidth requirements \agy{of the graphics engine\hjk{s}}. 
	\item \textbf{LLC\_Occupancy\_Tracer} \hjg{provides the number of \hjk{CPU} requests that are waiting for data to return from \agy{the} memory controller}. This counter indicates whether the \agy{CPU} cores are bandwidth limited.
	\item \textbf{LLC\_STALLS} counts the number of stalls due to a busy LLC. This indicates that the workload is \agy{main} memory latency limited.
	\item \textbf{IO\_RPQ} (IO \agy{R}ead \agy{P}ending \agy{Q}ueue occupancy) counts the stalls due to busy IOs. This counter indicates that the workload is IO limited.
\end{itemize}

To \emph{determine the threshold value} corresponding to \agy{each} performance counter, we \agy{perform} an offline phase that uses the results of the representative  \hjk{workload}s (from the performance counter selection phase). For these runs, we set a bound on the performance degradation \agy{(e.g., $1\%$)} when operating \hjk{in} MD-DVFS. We mark all the runs that have a performance degradation below this bound, and for the corresponding performance counter values\hjl{,} we calculate the \emph{mean} ($\mu$) and the \emph{standard deviation} ($\sigma$). \agy{We set the} threshold \agy{for each performance counter } \hjg{as $Threshold = \mu +  \sigma $ \cite{reimann2005background}.}

Our \emph{prediction algorithm uses} the performance counters to predict if the performance degradation of the running workload is less than a bound (e.g., $1\%$) when operating \hjk{in} MD-DVFS. To do so, the algorithm compares \hjk{the value of}  each of the selected performance counters to its threshold. If the  counter value  is greater than its threshold\hjk{,} then the algorithm keep\hjk{s} the SoC \hjk{at} the current \hjk{DVFS} operating point\hjk{;} otherwise, the algorithm decides to move the IO and memory domains to \hjn{the} lower \hjk{DVFS} operating point.

Our  prediction algorithm \hjk{has} an \emph{accuracy} of  $97.7\%$, $94.2\%$\hjn{,} and $98.8\%$ for single thread\hjk{ed CPU}, multi-thread\hjk{ed CPU}\hjl{,} and graphics workloads, respectively \hjm{(see Fig. \ref{fig:predict_perf_impact})}. The prediction algorithm \hjk{has} no \emph{false positive} predictions. This means that, there \hjk{are} no predictions \hjm{where} the algorithm decides to move the SoC to a lower DVFS operating point while the actual performance degradation is more than the bound.




\subsection{Holistic Power Management Algorithm}




SysScale 
implement\hjl{s} a power distribution algorithm in the \agy{PMU} firmware \hjl{to manage multi-domain DVFS and redistribute power among the domains}. PMU executes this algorithm \agy{periodically} at a configurable time interval called \emph{evaluation interval} ($30ms$ by default). PMU samples the performance counters and CSRs (the configuration and status registers) \agy{multiple} times in an evaluation interval (e.g, every $1ms$) and \agy{uses} the average value of each counter in the power distribution algorithm.

PMU switches the system between \agy{different} performance levels (i.e., DVFS operating points)  based on the predicted performance demand. Here we show the  power distribution algorithm \agy{that} switches the SoC between two operating points: \agy{high-} and \agy{low-performance} operating point\agy{s}. The system move\agy{s} to \hjk{the} high\agy{-performance} operating point if \emph{any} of the following five conditions is satisfied.
\begin{enumerate}
\item The aggregated \emph{static demand} requires higher memory bandwidth than a predefined threshold (STATIC\_BW\_THR).
\item The graphics \agy{engines} are \agy{bandwidth} limited (GFX\_LLC\_Misses > GFX\_THR). 
\item The \agy{CPU} core\agy{s are bandwidth} limited (LLC\_Occupancy\_Tracer > Core\_THR).
\item Memory latency is \agy{a bottleneck} (LLC\_STALLS > LAT\_THR)
\item IO latency is \agy{a bottleneck} (IO\_RPQ > IO\_THR). 
\end{enumerate}

If \emph{none} of these five conditions \hjk{are} true, PMU moves the system to the low-performance operating point. 
\hjk{In} the general case, \hjm{where} we have more than two {\tech} operating points, the above algorithm \hjk{decides} between two adjacent operating point\hjk{s} with dedicate\hjk{d} thresholds. 

When the SoC moves to the low-performance operating point, the PMU  \textbf{reduce\agy{s} the power budgets of \hjk{the} IO and memory domains and increase\agy{s} the power budget of \agy{the} compute domain 
}~\cite{lempel20112nd,rotem2013power,rotem2012power,21_doweck2017inside}.




The compute domain power budget management algorithm \agy{(PBM)} distributes the received power budget between \agy{CPU} cores and the graphics \agy{engines} according to the characteristics of the workloads that run on the\agy{se units}. The additional power budget can increase the frequency of a thermally limited compute domain. PBM is designed to keep the average power \agy{consumption} of the compute domain within the allocated power budget.

\subsection{Interaction with \agy{CPU} and Graphics DVFS}
C\agy{PU c}ores and \agy{the} graphics \agy{engines} support independent DVFS \agy{mechanisms.}
\agy{DVFS states are known as P-states}~\cite{rotem2009multiple,haj2018power,gough2015cpu,HDC_intel}).
The OS and \agy{the} graphics \hjg{driver} normally initiate DVFS requests for \hjl{the} \agy{CPU} core\agy{s} and  graphics \agy{engines}, respectively.
\agy{Each} request \agy{is} \hjl{handled by} the PBM algorithm inside the PMU firmware. PBM attempts to obey the DVFS request within the compute domain power budget constraint. If the request violates the  power budget, then PBM demotes the request and places the requestor in \agy{a} safe lower frequency \agy{that} keep\agy{s} the domain power within budget.

\hjk{Compute domain} DVFS mechanisms operate independently and are \emph{not} directly tied to {\tech}. {\tech} interacts with \hjk{compute domain DVFS mechanisms} only via power budget management. When {\tech} \agy{redistributes} power budget from IO and memory domains to the  compute domain, \agy{PBM} initiates DVFS \agy{in} the compute domain to utilize the new budget and increase \agy{the} domain's frequencies.
Similarly, \hjk{when} {\tech} \hjl{reduces the}  power budget \hjl{of} \agy{the compute domain} (e.g., due to high \agy{IO or} memory demand), \agy{PBM} adjusts (if needed) the compute domain's DVFS to keep the domain within the new power budget. 

\section{Implementation and Hardware Cost}
{\tech}~requires the implementation of the three components  inside the \hjg{SoC}. 

First, {\tech}~requires the implementation of the four performance counters (\hjm{as we describe in}  Sec.~\ref{subsec:demand_prediction_mechanism}). The implementation of these counters in any  \hjg{SoC} is straightforward. 

Second, {\tech}~requires the implementation of hardware and firmware \hjg{power management} flow that enable\hjk{s} the transition from one frequency/voltage \hjg{operating} point to another (as \hjm{we} summarize in Fig. \ref{fig:flow}). 
To implement \hjg{this} flow, \agy{two} capabilities \agy{are needed}\hjg{:} (1)  \agy{the} interconnect should support block and drain, which enables a seamless transition between frequencies\agy{; and} (2) the flow should be able to configure the memory controller and DDRIO with the relevant MRC values for each frequency to support multiple optimized DRAM frequencies. 
The MRC \hjg{values} can be determined at reset time by performing  MRC calculations for every supported frequency and saving the values inside an SRAM. 
The values \nv{are} retrieved from the SRAM during the flow transition. 
To support MRC updates, we need to dedicate approximately $0.5KB$ of SRAM, which corresponds to less than $0.006\%$ of Intel Skylake's die area \cite{21_doweck2017inside}. 
Additional capabilities, such as PLL\hjg{/DLL} re-locking and voltage adjustment, are supported in many SoCs. 
\hjg{All} of these capabilities exist inside the Skylake processor. 

Finally, the actual \hjg{demand prediction and power management algorithms}
\nv{are} implemented inside the PMU firmware.\footnote{A \hjn{fully-}hardware implementation is \hjm{also} possible. \hjk{H}owever\hjk{,} \hjk{such} \hjk{power management} flows are normally error-prone and require post-silicon \hjk{tuning}. As such, we choose to implement most of the flow within the power-management firmware (\hjk{e.g.,} Pcode\cite{gough2015cpu}).} The additional firmware code to support this flow is \hjg{approximately} $0.6KB$ ---corresponding to less than $0.008\%$ of \hjg{Intel Skylake}'s die area \cite{21_doweck2017inside}.  

\noindent \textbf{{\tech} Transition Time Overhead}. The actual latency of {\tech} flow is less than $10{\mu}$s. The latency
\nv{has the following \hjk{components}:}
\hjk{1) voltage transitions (approximately $\pm100mV$) of SA\_V and IO\_V voltage regulators (approximately $2{\mu}$s with a voltage regulator slew rate of $50mV/{\mu}$s),}
2) draining IO interconnect \hjg{request} buffers \nv{(}less than $1{\mu}$s\nv{)}, 3) exiting DRAM self-refresh \nv{(}less than $5{\mu}$s with a fast training process\nv{)}, 4) loading the optimized DRAM configuration register from SRAM into configuration registers \nv{(}less than $1{\mu}$s\nv{)}, and 5) firmware latency and other flow overheads \nv{(}less than $1{\mu}$s\nv{)}.

\section{Evaluation Methodology}
\label{sec:method}


We evaluate SysScale \hjg{using} real systems that employ \hjg{Intel} Broadwell~\cite{5_nalamalpu2015broadwell} and \hjg{Intel} Skylake~\cite{21_doweck2017inside} \hjg{SoCs}. We use two distinct methodologies for 1) collecting motivational data and 2) evaluating SysScale. The reason is that we \hjg{would like}  to demonstrate the potential benefits of SysScale on \hjg{the} previous generation SoC (i.e., Broadwell) of our target Skylake SoC\hjg{,} before we implement it in the  Skylake SoC. 

\noindent \textbf{Methodology for Collecting Motivational Data.}
To collect motivational data, we use a Broadwell-based system on which we emulate \hjk{a crude version of} SysScale's static behavior, i.e., the \hjg{multi-domain} DVFS setup (MD-DVFS) that we use in Sec. \ref{sec:motiv}. We use \hjg{three} steps to attain MD-DVFS setup.
First, we reduce the DRAM frequency (to $1.06GHz$) using \hjg{the} BIOS settings. \hjk{Doing so} boots the system with optimized MRC values for \hjg{the} DRAM interface (\hjm{as we} explain in Sec. \ref{sec:mrc})\hjk{,} which exist in the BIOS. 
Second, since \hjk{the} memory controller and IO interconnect share the same voltage (V\_SA \circled{1} in Fig. \ref{skl_arch}), we also proportionally reduce the IO interconnect clock frequency (to $0.4GHz$) to align the voltage levels of \hjg{the} IO interconnect and memory controller. 
\hjg{Third,} we reduce the shared voltage \hjg{and DDRIO voltage} (i.e., V\_SA and V\_IO)  by approximately $20\%$ and $15\%$, \hjg{respectively,} which we determine based on the voltage/frequency curves of all components that share th\hjk{ese} voltage\hjk{s}. To configure the voltage and frequency to the new values, we use \hjg{the} In-Target Probe (ITP) hardware debugger unit\hjm{, which we explain below}.

\noindent \textbf{Methodology for  Evaluating SysScale.}
\hjk{We} implement \hjg{\tech} on \hjk{the} \hjg{Intel} Skylake SoC \cite{21_doweck2017inside}. Table~\ref{tbl:sys_setup} shows the \hjg{major system} parameters. \hjg{For our baseline measurements} we disable SysScale on \hjk{the} \hjg{same} SoC.

\begin{table}[h]
\centering
\caption{SoC and memory \hjg{parameters} 
}
\label{tbl:sys_setup}
\begin{tabular}{cl}
\hline
\multirow{3}{*}{SoCs}                 & \multicolumn{1}{c}{M-5Y71\cite{intel5Y71} Broadwell microarchitecture.}                                                                                             \\
                                            & M-6Y75\cite{intel6Y75} Skylake microarchitecture.                                                                                                                   \\
                                            & \begin{tabular}[c]{@{}l@{}}CPU Core Base Frequency: 1.2GHz\\
                                    Graphics Engine Base Frequency: 300MHz\\        
                                    L3 cache (LLC): 4MB.\\ Thermal Design Point (TDP): 4.5W\\ Process technology \hjg{node}: 14nm\end{tabular} \\ \hline
\multicolumn{1}{l}{\multirow{2}{*}{Memory}} & LPDDR3-1600MHz \cite{no2013jesd79}, non-ECC,                                                                                                                          \\
\multicolumn{1}{l}{}                        & dual-channel, 8GB capacity                                                                                                                                  \\ \hline
\end{tabular}
\end{table}


\noindent \textbf{In-Target Probe (ITP).} ITP is a silicon debugger tool that connects to an Intel SoC through \hjg{the} JTAG port~\cite{williams2009low}. We use ITP 1) to set breakpoints at which the SoC halts when a specified event occurs and 2) configure \hjg{the} SoC's control and status registers (CSRs) and model specific registers (MSRs)~~\cite{intel_itp}.

\noindent \textbf{Power Measurements.} We measure power \hjk{consumption} by using a National Instruments Data Acquisition (NI-DAQ) card (NI-PCIe-6376~\cite{NIDAQ}), whose sampling rate is up to 3.5 \hjg{M}ega-samples-per-second (MS/s). Differential cables transfer multiple signals from the power supply lines on the motherboard to the \hjg{NI-}DAQ card in the host computer that collects the power measurements. By using NI-DAQ, we measure \hjg{power on} up to 8 channels \hjg{simultaneously}. We connect \hjg{each} measurement channel to one  voltage regulator of \hjg{the SoC}. The power measurement  accuracy of the NI-PCIe-6376 is $99.94\%$.

\noindent \textbf{Workloads.} We evaluate {\tech} with three classes of workloads that are widely used for evaluating mobile SoCs: 1) To evaluate  CPU core performance, we use  \hjg{the} SPEC CPU2006 benchmarks~\cite{SPEC}. We use \hjg{the SPEC CPU2006 benchmark} score as \hjg{the} performance metric. 2) To evaluate  computer graphics performance, we use \hjg{the} 3DMARK benchmarks~\cite{17_3DMARK}. We use frames per second (FPS) as \hjg{the} performance metric. 3) To evaluate the \hjg{effect} on battery life, we use \hjg{a} set of workloads that are typically used to evaluate the battery life of mobile devices such as, web browsing, light gaming, video conferencing\hjm{,} and video playback \cite{19_MSFT}. We use average power consumption as \hjg{our battery life} evaluation metric. 

\noindent \textbf{Methodology for  Comparison  to Prior Works.}
We compare {\tech} to the two most relevant prior works, MemScale \cite{deng2011memscale} and CoScale \cite{deng2012coscale}. Unfortunately, 1) there is no real system available that implements these techniques , and 2) MemScale and CoScale techniques save \emph{only} average power consumption 
\hjg{without having the ability to redistribute excess power to other domains. In other words, they cannot boost the performance of the compute domain when applying DVFS to \hjk{the} memory domain.}
Therefore, \hjg{to make our comparison useful}, we \emph{assume} that MemScale and CoScale can redistribute their saved power \hjk{in the memory domain} to increase \hjg{the} compute domain power budget (similar to \tech). \hjg{We call these improved versions of MemScale and CoScale, MemScale-Redist and CoScale-Redist, respectively \hjk{(sometimes abbreviate\hjm{d as} MemScale-R and CoScale-R)}.} 

We project  the performance improvements of MemScale and CoScale compared to our baseline using three steps. 
First, we estimate the average power savings \hjg{of MemScale and CoScale} by using per-component measurements that we carry out on \hjg{our} Skylake system using power measurement tools \hjg{that} we \hjg{describe} in this section.
Second, we build a performance/power model that maps \hjg{an} increase in the power budget of \hjk{the} compute domain to an increase in the CPU core or the graphics engine frequencies. For example, the model can show that a $100mW$ increase in compute power budget can lead to an increase in the core frequency by $100MHz$. To build \hjg{this} model, we use performance and power measurements that we carry out on our Skylake system \hjg{using} multiple CPU core and graphics engine \hjg{frequencies}.
Third, we use the \emph{performance scalability}\footnote{We define performance scalability of a workload with respect to CPU frequency as the \hjk{performance} improvement the workload \hjm{experiences} with \hjm{unit} increase \hjm{in}  frequency.} of the running workload with CPU frequency to project the actual performance improvement of the workloads \hjg{with MemScale-Redist and CoScale-Redist}.

\section{Results}\label{sec:results}

We present performance and average power benefits\footnote{\hjk{E}nergy efficiency, \hjm{i.e.,} energy-delay-product (EDP) \hjn{\cite{gonzalez1996energy}}, improves  proportionally to \hjk{performance or} average power \hjk{consumption}, 
since {\tech} improves \hjm{either} \hjk{performance} \hjm{within a fixed power budget or} average power \hjm{consumption within a fixed} performance requirement.} obtained with {\tech} when it is \hjk{implemented} in an Intel Skylake SoC compared to the same SoC with {\tech} disabled. We also compare {\tech} to two prior works MemScale~\cite{deng2011memscale} and CoScale~\cite{deng2012coscale}.  {\tech}\hjl{'s} performance and average power benefits are \emph{measured} results, while the results for MemScale-Redist and CoScale-Redist are \emph{projected}, as we explain in Sec. \ref{sec:method}. 
We \hjl{evaluate} three workload categories: CPU (Sec. \ref{res:cpu}), graphics (Sec. \ref{res:graphics}), and battery life workloads (Sec. \ref{res:battery}). 
We also analyze  sensitivity to different SoC thermal-design-power (TDP) levels and DRAM frequencies (Sec. \ref{res:sens}).

\subsection{Evaluation of CPU Workloads}
\label{res:cpu}

 \begin{figure*}[!ht]
  \begin{center}
  \includegraphics[trim=0.7cm 0.9cm 0.7cm 1.2cm, clip=true,width=0.95\linewidth,keepaspectratio]{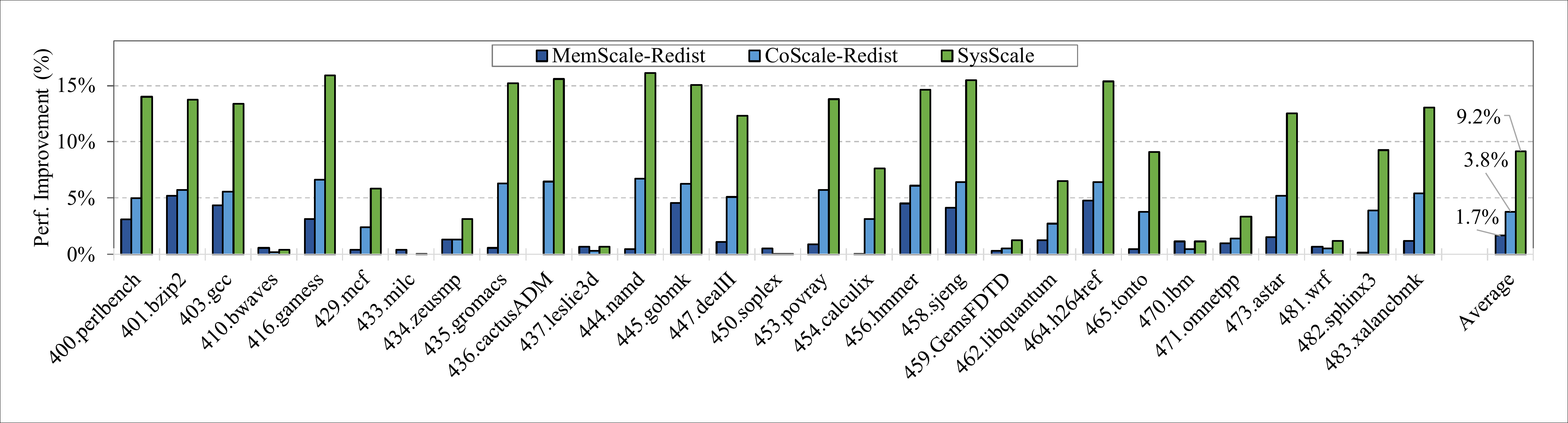}\\
  \caption{Performance improvement of  MemScale-Redist, CoScale-Redist\hjk{,} and SysScale \hjk{on} SPEC CPU2006 \hjk{workloads}.}\label{fig:core_res}
  \end{center}
 \end{figure*}

Fig. \ref{fig:core_res} reports the performance improvements of  MemScale-R, CoScale-R, and {\tech}~over our baseline system.
We make four key observations. 

First, {\tech} improves real system performance by $9.2\%$ on average. This result is significant as it is obtained on a real \hjm{system}. 

\hjk{Second,} {\tech} provides  $5.4 \times$ and $2.4\times$  \hjm{the} performance improvement \hjm{of} MemScale-R and CoScale-R, \hjk{respectively}.
This is because 1) SysScale is holistic, taking into account all SoC domains (and components in each domain), whereas MemScale and CoScale  consider only memory DVFS and coordinated (CPU and memory subsystem) DVFS, respectively, and 2) MemScale and CoScale do not dynamically optimize \hjk{the} DRAM interface configuration registers and voltage after each DVFS change.
\hjk{Also recall} that the performance improvement of MemScale-R and CoScale-R is projected based on the estimated average power reduction of each  technique \hjl{(Sec. \ref{sec:method})}, whereas {\tech} improvements are real measured improvements. 

Third, the performance benefit of {\tech}  \hjk{correlates with} the \emph{performance scalability}
of the running workload with CPU frequency. Highly-scalable workloads (i.e., those bottlenecked by CPU cores, \hjk{such as \emph{416.gamess} and \emph{444.namd}}) have the highest performance gains. In contrast, workloads that are heavily bottlenecked by main memory, such as \emph{410.bwaves} and \emph{433.milc}, have almost no performance gain.
We note that a highly scalable workload (e.g., \emph{416.gamess}) benefits in two ways from {\tech}:
1) because the workload is not bottlenecked by main memory, reducing memory frequency reduces power consumption without affecting performance,
2) because the workload is bottlenecked by CPU performance, redistributing the power budget from memory and IO to the CPU increases the workload’s performance.


\hjk{Fourth, i}f a workload has different execution phases bottlenecked by different SoC domains (e.g., compute versus memory), then  {\tech} dynamically adjusts the frequencies of the multiple domains to improve system performance. For example, \emph{473.astar} has execution phases of up to several seconds of low memory bandwidth demand (e.g., ${\sim}1GB/s$) and high memory bandwidth demand (about ${\sim}10GB/s$)\hjk{,} as illustrated \hjk{in} Fig. \ref{fig:mem_bw}(a)\hjk{, and {\tech} significantly improves its performance by \hjk{$13\%$}.}

We conclude that {\tech} significantly improves CPU core performance by holistically applying DVFS to SoC domains based on dynamic performance demands and dynamically redistributing the \hjk{excess} power budget between domains, while carefully optimizing the memory interface during DVFS transitions. 

\subsection{Evaluation of Graphics Workloads}
\label{res:graphics}

Typically, the performance of a graphics  workload  is highly scalable with \hjl{the} graphics engine frequency. When running graphics workloads, the power budget management algorithm (PBM \cite{lempel20112nd,rotem2015intel}) of the PMU normally allocates only $10\%$ to $20\%$ of the compute domain power budget to the CPU cores, while the graphics engines consumes the rest of the power budget~\cite{9_rotem2011power,rotem2012power,rotem2013power}. 
For a mobile system, while running a graphics workload\hjk{,} the CPU cores normally run at the \hjk{most} energy efficient frequency $Pn$~\cite{haj2018power} (i.e., the maximum possible frequency at the minimum functional voltage ($V_{min}$)). Moreover, at a very low TDP (e.g., $3.5W$ and $4.5W$), the effective CPU frequency is reduced below \textit{Pn} by using hardware duty cycling (HDC\footnote{Hardware Duty Cycling, also known as SoC Duty Cycling, implements coarse grained duty cycling by using C-states with power gating in contrast to T-states that \agy{use} clock gating \cite{HDC_intel}.})~\cite{HDC_intel}. 

\begin{figure}[!h]
  \begin{center}
  \includegraphics[trim=.99cm .8cm .8cm .8cm, clip=true,width=\linewidth,,keepaspectratio]{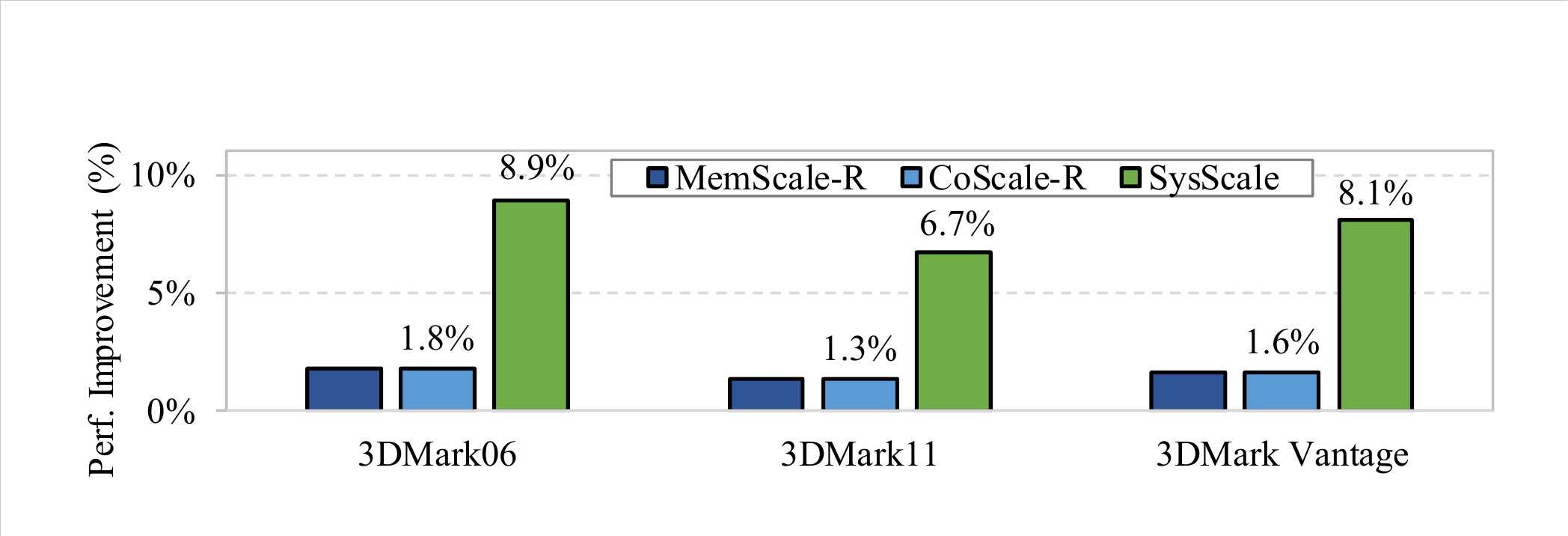}\\
  \caption{Performance improvement of MemScale\-R, CoScale-R, and {\tech} \hjk{on} computer graphics workloads.}
  \label{fig:gfx_res}
  \end{center}
 \end{figure}

Fig. \ref{fig:gfx_res} shows the performance improvement of MemScale-R, CoScale-R\hjk{, and} {\tech}\hjl{,}  
compared to our baseline when running three different 3DMark \cite{17_3DMARK} graphics workloads. 
We make two key observations.
First, {\tech} improves the system performance of 3DMark06, 3DMark11\hjn{,} and 3DMark Vantage  by $8.9\%$, $6.7\%$ and  $8.1\%$, respectively. 
{\tech} improves performance because it boosts the graphics engines\hjl{,} frequency by
 redistributing  the power budget across
 three SoC domains. 


Second, {\tech} provides \hjk{approximately} $5\times$ the performance improvement of MemScale-R and CoScale-R.
MemScale-R and CoScal\agy{e}-R have similar performance improvements because the average power savings of these two techniques is identical. The reason is that\hjk{,} in these workloads\hjk{,} the CPU cores \agy{run} at the lowest possible frequency, \hjk{and} therefore CoScale (which applies DVFS to CPU cores in addition to the memory subsystem) cannot further scale down the CPU  frequency. 


We conclude that the saved power budget from IO and memory domains can be used to raise the frequency of the graphics engines and improve graphics performance. 

\subsection{\hjm{Evaluation of} Battery \hjk{Life Workloads}}
\label{res:battery}

Unlike  CPU  and graphics workloads that always benefit from higher performance, the battery life workloads 
have two characteristics: 1) \hjk{t}heir performance demands are fixed\hjk{;} for example\hjk{,} in video playback running at 60 frames per second, \hjk{each} frame need\hjm{s} to \hjk{be} processed and displayed on the display panel within \hjk{16.67} milliseconds,
and 2) they have long idle phases where the system enters into idle power states (C-states~\cite{haj2018power,acpi,gough2015cpu,haj2020techniques}). We note that according to our measurements, the active state (i.e., $C0$ power state) residency of these workloads is between $10\%$-$40\%$, while the SoC is in idle state (e.g., C2, C6, C7, or C8 \cite{haj2018power,acpi,gough2015cpu,haj2020techniques}) during the remaining execution time of the workload. In the $C0$ power state, typically the compute domain (i.e., CPU cores and graphics engines) operates in the lowest possible frequencies, while in all other power states\hjn{,} the compute domain is idle (\hjk{i.e.,} clock- or power-gated).

\begin{figure}[!h]
  \begin{center}
  \includegraphics[trim=.8cm .8cm .8cm 0.8cm, clip=true,width=1.0\linewidth,keepaspectratio]{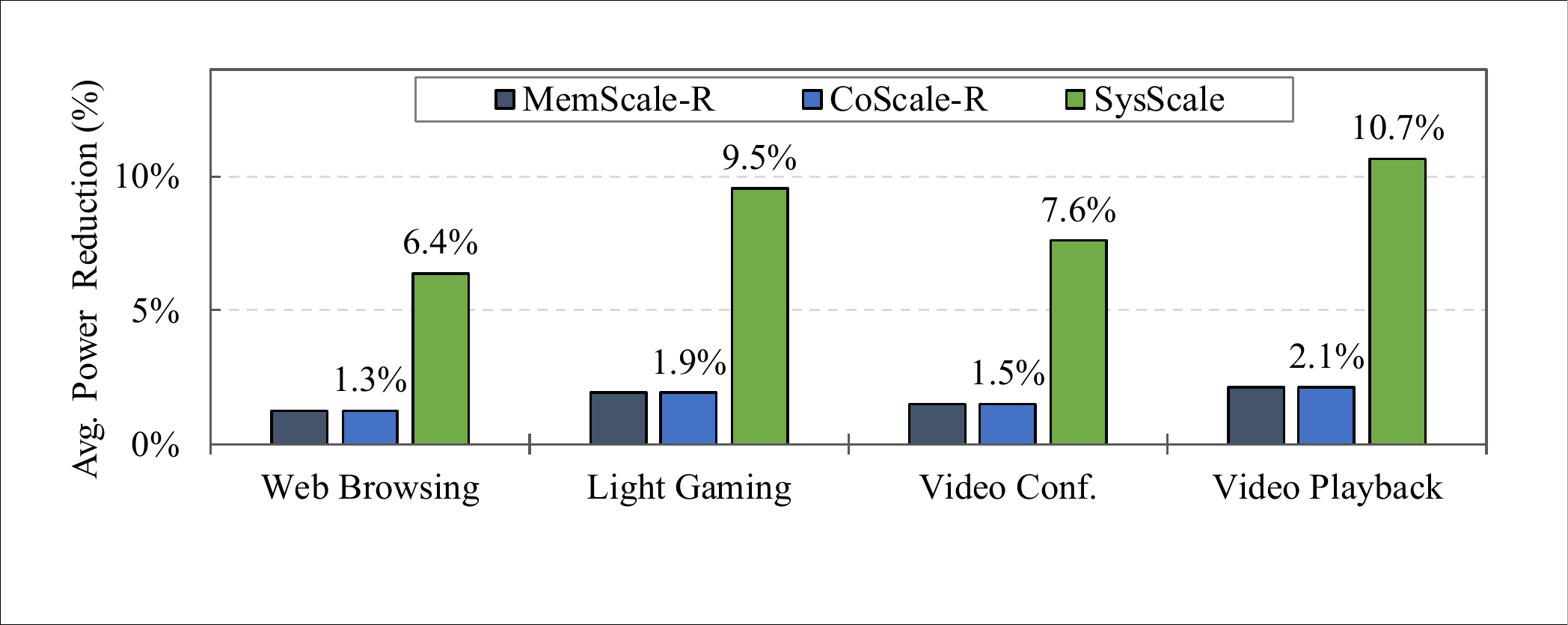}\\
  \caption{\hjk{Average power} reduction of \hjm{MemScale\hjg{-R}, CoScale\hjg{-R}, and {\tech} on} representative battery life workloads. }\label{fig:avg_power_res}
  \end{center}
 \end{figure}

Fig. \ref{fig:avg_power_res} shows the \hjk{SoC}  average power reduction when running four \hjk{representative} battery life workloads \cite{19_MSFT}, web-browsing, light-gaming, video conferencing\hjm{,} and video-playback, \hjk{with} a single HD display panel (e.g., the laptop display) is active. 
We make three key observations.

First, {\tech} reduces average power consumption of web browsing, light gaming, video conferencing\hjn{,} and video playback workloads by $6.4\%$, $9.5\%$, $7.6\%$\hjm{,} and $10.7\%$, respectively\hjg{, on our real system}.

Second, {\tech} provides \hjk{approximately} $5\times$  the power reduction of MemScale-R and CoScale-R.
MemScale-R and CoScale-R provide similar average power reduction benefits to each other, \agy{since} in
battery life workloads, the CPU cores \agy{run} at the lowest possible frequency, and therefore CoScale cannot further reduce the CPU cores\hjl{,} frequency.

Third, {\tech} provides significant power savings for battery life workloads for all SoC power states in which DRAM is active (i.e., not in self-refresh).
For example, in the video playback workload, the SoC transitions between three power states during the processing of each video frame: $C0$, $C2$\hjm{,} and $C8$ with residencies of $10\%$, $5\%$\hjm{,} and $85\%$, respectively (not graphed). \hjl{DRAM is active} only in $C0$ and $C2$ power states, while in C8\hjk{,} DRAM is in self-refresh.
Therefore, {\tech}  applies DVFS to the IO and memory domains \hjl{only} during $C0$ and $C2$ \hjk{states} in \hjk{the} video playback workload. 

We conclude that for battery life  workloads\hjk{,} which have fixed performance requirements, applying {\tech} \hjk{as a} holistic DVFS mechanism for \hjk{all three SoC} domains significantly reduces the average \hjk{SoC} power consumption\hjk{.}

\subsection{Sensitivity Analysis}
\label{res:sens}
In this section, we investigate the effect of system parameters on {\tech} performance and power gains. 

\noindent \textbf{SoC Thermal Design Point (TDP).} 
The evaluation results presented in Sections \ref{res:cpu}, \ref{res:graphics}, and \ref{res:battery} are for an SoC (i.e., Skylake M-6Y75 \cite{intel6Y75}) with a TDP of $4.5W$, as we show in Table \ref{tbl:sys_setup}. This SoC has a configurable TDP ranging from $3.5W$ up to $7W$. The Skylake architecture itself can scale up to a TDP of $91W$ \cite{91w_intel} for a high-end desktop.

\begin{figure}[!h]
  \begin{center}
  \includegraphics[trim=.7cm .7cm .7cm .7cm, clip=true,width=1.0\linewidth,keepaspectratio]{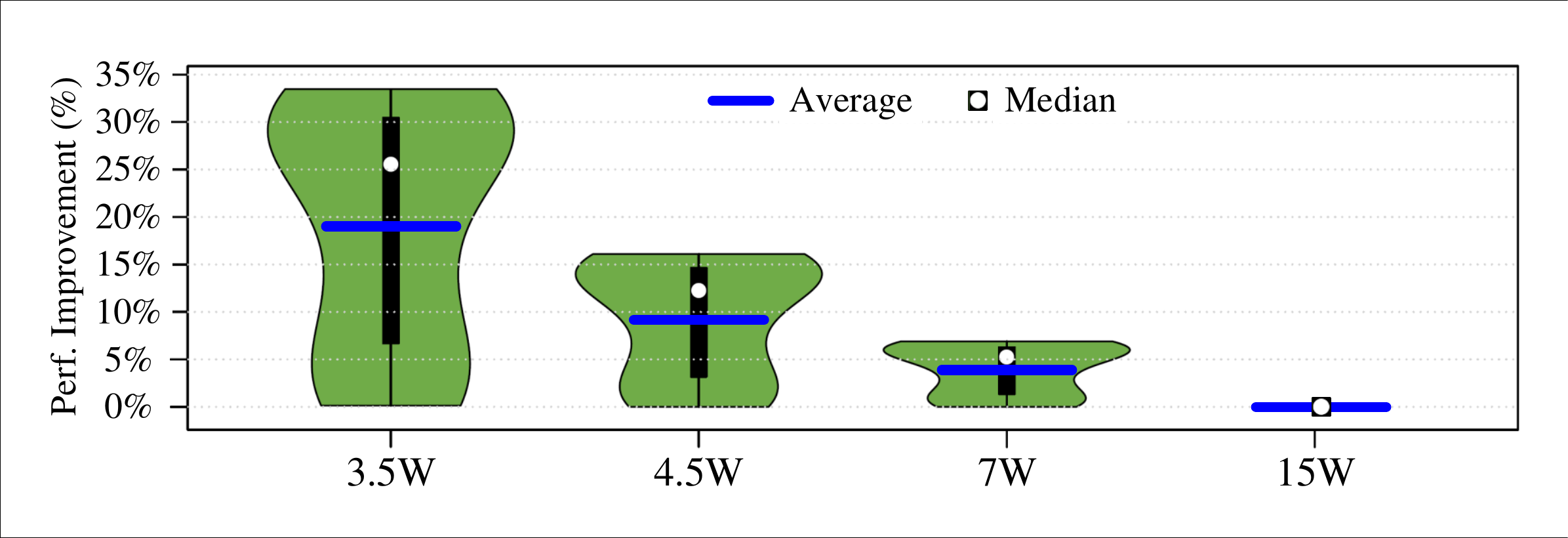}\\
  \caption{{\tech} performance benefit vs. TDP on SPEC CPU2006 workloads. 
  }\label{fig:sens_study}
  \end{center}
 \end{figure}

Fig. \ref{fig:sens_study} shows the average performance improvement on SPEC \hjm{CPU}2006 \hjm{workloads} when running {\tech} on systems with different TDPs.
 \hjm{Violin plots show the  distribution and density of performance improvement points across different workloads compared to each TDP's baseline.}
We make two key observations.
First, at a constrained TDP of 3.5W, {\tech} improves performance by \hjk{up to \hjk{$33\%$} (}$19.1\%$ \hjk{on average)}.
Second, as TDP increases, {\tech}’s performance benefit reduces. This is because power becomes ample and  there is less need to redistribute \hjk{it} across domains.

We also evaluate battery life workloads and found that {\tech}'s average power savings are not affected by the TDP. In these workloads,  the compute domain typically operates at the lowest  possible frequencies for the CPU cores and graphics engines  (i.e., the $Pn$ P-state ~\cite{haj2018power,gough2015cpu}) regardless of the TDP of the SoC.

We conclude that {\tech} significantly improves the performance of especially \hjk{TDP-}constrained SoCs while it improves  battery life (energy consumption) across the entire TDP range (e.g., $3.5W$--$91W$ \cite{intel6Y75,91w_intel}).

\noindent \textbf{More DRAM Frequencies.}
In our previous experimental results, we use a system with a base DRAM frequency of $1.6GHz$. 
We \hjk{also} evaluate a system with different DRAM devices and frequencies. Based on our evaluation of {\tech} with different DRAM types and frequencies, we make three key observations.

First, {\tech} supports other DRAM devices with different DRAM frequencies. 
The average power savings (freed-up power budget that is dynamically utilized to boost \hjk{the compute} domain) obtained by {\tech} when scaling DDR4's 
operating point from $1.86GHz$ to $1.33GHz$ is \hjk{approximately $7\%$} lower than \hjk{that} when scaling \hjk{LP}DDR3 
operating point 
from $1.6GHz$ down to $1.06GHz$. 

Second, {\tech} supports multiple operating points for the same DRAM device. For example, we could use $0.8GHz$ as the lowest DVFS operating point of the memory subsystem. However, we observe that the $0.8GHz$ operating point is not energy efficient when compared to $1.06GHz$ due to two main reasons. 1) \hjg{T}he system agent voltage (V\_SA) already reaches the minimum functional voltage ($V_{min}$) when scaling DDR frequency to $1.06GHz$. Therefore, reducing the DRAM  frequency beyond $1.06GHz$ does not provide significant benefits as the V\_SA voltage remains the same.
2) \hjg{W}hen comparing Figures \ref{fig:predict_perf_impact}(a,d,g) and Figures \ref{fig:predict_perf_impact}(b,e,h), we observe that the average performance degradation of the traces when reducing the DRAM frequency from $1.6GHz$ down to $0.8GHz$ is $2\times$-$3\times$ higher than that when reducing it from $1.6GHz$ down to $1.06GHz$. As such, we implement only two operating points (i.e., $1.6GHz$ and $1.06GHz$) \hjk{in} our real system.

Third, a \textbf{finer\hjg{-}grained} DVFS of the three SoC domains can increase the benefits of {\tech}. For instance, supporting additional DDR frequencies between the $1.6GHz$ to $0.8GHz$ range can increase the performance and/or the average power of workloads with memory demand that fall in between existing frequencies. Unfortunately, current commercial DDR devices support only few frequencies (e.g., \hjk{LP}DDR3 supports only $1.6GHz$, $1.06GHz$, and $0.8GHz$). 

We conclude that the benefits of {\tech} can be higher \hjl{with} more control over the DRAM frequencies. To achieve this, we recommend that the \textbf{DRAM vendors} enable finer-grained control on DDR frequency in future generations to \hjk{help} \hjl{increase} the energy efficiency of modern mobile SoCs. 

 



\section{Related Work}
To our knowledge, this is the first work to 
1) \hjg{enable coordinated and highly\ch{-}efficient DVFS across all SoC domains to increase the energy efficiency of mobile SoCs and} 
2) provide \hjg{multi-domain} DVFS results from real \hjk{a} \hjg{modern mobile SoC (Intel Skylake)} running  real mobile system benchmarks. 

\noindent {\textbf{Compute Domain DVFS.}}
Many prior works \ch{propose} DVFS \ch{for the} compute domain \ch{(i.e.,} the CPU cores and graphics engines) \cite{mobius2013power, kolpe2011enabling, hanumaiah2012energy, kim2008system, herbert2007analysis, miyoshi2002critical, sundriyal2018core,rotem2009multiple,haj2018power,gough2015cpu,HDC_intel,zhang2016maximizing,imes2018handing,isci2006live,isci2006analysis,isci2005long}.
\ch{DVFS operating points for the compute domain, known as P-states~\cite{rotem2009multiple,haj2018power,gough2015cpu,HDC_intel}, are typically managed by the OS and the graphics drivers \hjk{for} the CPU \hjk{cores} and graphics engines, respectively}.
\ch{These mechanisms optimize voltage and frequency assuming a fixed power budget for the compute domain. Unlike {\tech}, they do not enable optimization and distribution of power \hjg{budget} across different \hjk{SoC} domains and thus provide limited energy efficiency benefits in mobile SoCs. 
}

\noindent {\textbf{Memory DVFS and Coordinated DVFS.}}
Recent works in \hjg{memory} DVFS 
\cite{david2011memory,deng2011memscale,chen2011predictive,deng2012coscale,felter2005performance,li2007cross,zhang2016maximizing,imes2018handing,deng2012multiscale,chang2017understanding} for modern 
SoCs focus only on improving energy efficiency of
\ch{the} memory domain (e.g., MemDVFS \cite{david2011memory}\hjg{, Voltron \cite{chang2017understanding}}, and MemScale \cite{deng2011memscale}) or limited components of two domains (e.g., CoScale \cite{deng2012coscale} and other works \cite{chen2011predictive,felter2005performance,li2007cross}). 

\hjg{In Sections \ref{res:cpu}, \ref{res:graphics}\ch{,} and \ref{res:battery}, we qualitatively
and quantitatively compare {\tech} to two closely
related prior works, MemScale~\cite{deng2011memscale} and CoScale~\cite{deng2012coscale}\ch{.}  
We show that {\tech} significantly \ch{increases} system performance and \ch{reduces} energy consumption compared to the two mechanisms \hjk{due to two reasons:} 
1) SysScale is holistic, taking into account all SoC domains (and components in each domain), whereas MemScale and CoScale consider only memory DVFS and\ch{/or compute domain} DVFS 
and 2) MemScale and CoScale do not dynamically  optimize DRAM interface configuration registers and voltage after each DVFS change.}

Current \hjg{memory DVFS} approaches \hjg{\cite{chang2017understanding,david2011memory,deng2011memscale,chen2011predictive,deng2012coscale,felter2005performance,li2007cross,zhang2016maximizing,imes2018handing}} \hjk{all} have
\hjg{ one or more of} \ch{three main} drawbacks that make them \hjg{inefficient for} modern mobile SoCs.
First, \hjk{many of} \ch{these} prior works 
are not \hjk{directly} \hjg{optimized} for modern mobile SoC architectures that integrate \hjk{\emph{many}} \hjg{components} into the three SoC domains 
in a \hjk{very} thermally \hjk{constrained} environment. \hjk{Many works} target server architectures, which have different structure, constraints\ch{,} and workloads \hjg{than in mobile systems}.     
Second, \hjk{past works} scale \textit{\ch{only}} \hjg{one or two} \hjk{system} domains at a time. \hjm{Such limited} scaling  \hjg{provides small} benefits for mobile SoCs (as we show in \hjm{Sections \ref{sec:motiv} and} \ref{sec:results}) and\hjk{,} when accounting for the relatively high \hjk{DVFS} transition costs, the benefits diminish even further. 
Third, \hjk{past works} 
\hjg{do not dynamically optimize} \hjk{the} DRAM interface configuration registers \cite{mrc1,mrc2,mrc3}, which degrades performance and may negate potential gains (as we show in \hjm{Sections \ref{sec:motiv} and} \ref{sec:results}).

\section{Conclusion}


We propose {\tech}, the first work to enable coordinated and highly-efficient DVFS across all SoC domains to increase the energy efficiency of mobile SoCs. {\tech} introduces the ability to \ch{optimize and efficiently} redistribute the total power budget across all SoC domains according to the performance demands of each domain.
{\tech} is implemented in the Intel Skylake SoC for mobile devices.
We show that it significantly improves the performance of real CPU and graphics workloads (by up to $16\%$ and $8.9\%$, respectively, for $4.5W$ TDP) and reduces the average power consumption of battery life workloads (by up to $10.7\%$) on \ch{a} real Intel Skylake system. 
We conclude that {\tech} is an effective approach to balance power consumption \hjg{and performance} demands  \hjg{across all SoC domains} in a sophisticated heterogeneous \hjk{mobile} SoC to improve energy efficiency and performance.


\section*{Acknowledgments} We thank the anonymous reviewers of \hjg{ISCA \hjk{2020}} for feedback and the SAFARI group members for
feedback and the stimulating intellectual environment they provide. 
\\
\\



%


\SetTracking
 [ no ligatures = {f},
 outer kerning = {*,*} ]
 { encoding = * }
 { -40 } 

{

  \let\OLDthebibliography\thebibliography
  \renewcommand\thebibliography[1]{
    \OLDthebibliography{#1}
    \setlength{\parskip}{0pt}
    \setlength{\itemsep}{0pt}
  }
  \bibliographystyle{IEEEtranS}
  \bibliography{refs}
}

\end{document}